\documentclass{mn2e}
\usepackage{epsfig}

\title[Near-IR Spectroscopy of $\eta$ Carinae]{Dissecting the
Homunculus Nebula Around Eta Carinae With Spatially Resolved
Near-Infrared Spectroscopy}

\author[N.\ Smith]{Nathan Smith\thanks{Email: \tt
nathans@casa.colorado.edu}\thanks{Visiting Astronomer, Cerro Tololo
Inter-American Observatory, National Optical Astronomy Observatories,
operated by the Association of Universities for Research in Astronomy,
Inc., under cooperative agreement with the National Science
Foundation.} \\ \noindent Center for Astrophysics and Space Astronomy,
University of Colorado, 389 UCB, Boulder, CO 80309, USA}

\date{Accepted 2002 August 11. Received 2002 August 5; in original form 2002 July 10}
\volume{337}
\pagerange{1252--1268}
\pubyear{December 21, 2002}

\def\arcdeg{\degr}

\def\gtrsim{\ga}
\def\micron{\mu{\rm m}}

\begin{document}
\label{firstpage}
\maketitle
\begin{abstract}

Near-infrared emission lines provide unique diagnostics of the
geometry, structure, kinematics, and excitation of $\eta$ Carinae's
circumstellar ejecta, and give clues to the nature of its wind. The
infrared spectrum is a strong function of position in $\eta$ Car's
nebula, with a mix of intrinsic and reflected emission.  Molecular
hydrogen traces cool gas and dust in the polar lobes, while [Fe~{\sc
ii}] blankets their inner surfaces.  These lines reveal the back wall
of the SE polar lobe for the first time, and give the clearest picture
yet of the 3-D geometry.  Additionally, collisionally-excited [Fe~{\sc
ii}] reveals the kinematic structure of a recently discovered `Little
Homunculus' expanding inside the larger one.  Equatorial gas in the
`Fan', on the other hand, shows a spectrum indicating recombination
and fluorescent Ly$\alpha$ pumping.  Some equatorial ejecta glow in
the He~{\sc i} $\lambda$10830 line, showing evidence for material
ejected in the 1890 outburst of $\eta$ Car.  Closer to the star, the
compact `Weigelt blobs' are marginally resolved, allowing their
infrared spectrum to be separated from the star for the first time. In
general, infrared spectra reveal a coherent, directional dependence of
excitation in the Homunculus: polar ejecta are collisionally excited,
whereas equatorial ejecta are dominated by fluorescence and normal
photoexcitation.  These are important clues to the geometry of the
central star's UV radiation field. Reflected near-infrared emission
lines also reveal interesting latitudinal dependence in the stellar
wind.

\end{abstract}


\begin{figure*}\begin{center}
\epsfig{file=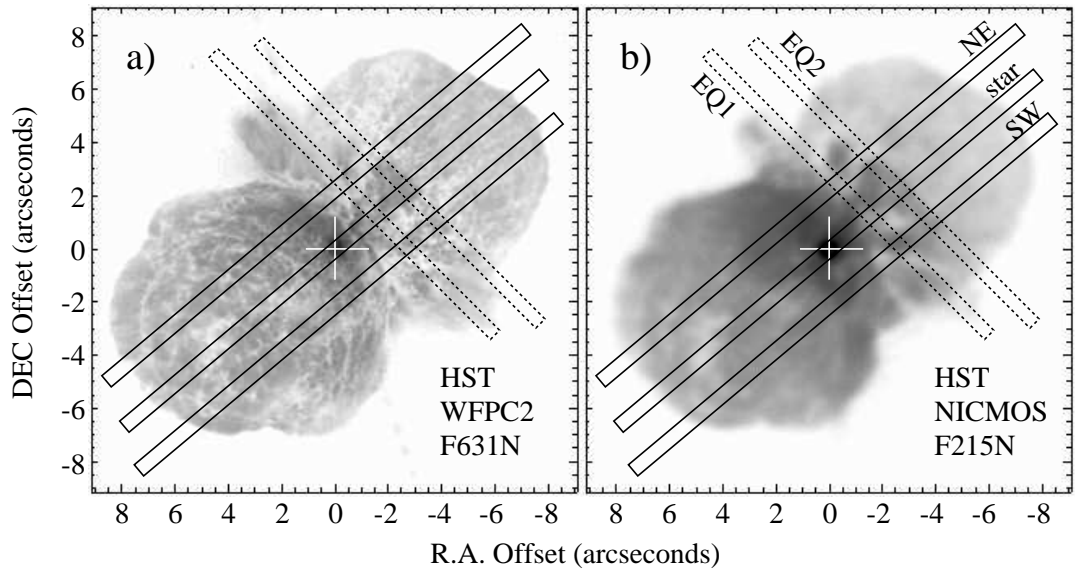,height=3in}\end{center}
\caption{(a) A red {\it HST}/WFPC2 image of $\eta$ Car (see Morse et
al.\ 1998) showing orientations of OSIRIS apertures.  (b) same as $a$
but with a $\sim$2 $\micron$ continuum {\it HST}/NICMOS image from
Smith \& Gehrz (2000).  Dotted boxes show positions of long-slit
apertures crossing equatorial ejecta, discussed in \S 5.  Slit
positions are labeled with designations referred to in the text.}
\end{figure*}

\section{INTRODUCTION}

The idiosyncratic massive star $\eta$ Carinae and its bipolar
Homunculus Nebula provide spectacular examples of a dense stellar wind
and processed ejecta atop the HR Diagram.  Its proximity and
brightness allow detailed multiwavelength study, offering insight to
the limits nature imposes on the stability of stars.  $\eta$ Car
exhibits extreme spatial, temporal, and spectral complexity throughout
the electromagnetic spectrum; infrared (IR) wavelengths afford us no
exception, but they provide unique diagnostics of the wind and ejecta.

The shape, orientation, kinematics, and excitation of the Homunculus
offer the best ways to investigate ejection physics of the Great
Eruption, since most material was ejected in the 1840's (Morse et al.\
2001; Smith \& Gehrz 1998; Currie et al.\ 1996).  The Homunculus is a
dusty nebula dominated by reflected starlight at optical wavelengths.
This makes it problematic to study the kinematics and excitation of
the Homunculus itself, because narrow emission lines originating in
the polar lobes overlap with reflected emission (e.g., Hillier \&
Allen 1992; Davidson et al.\ 2001).  IR wavelengths offer a way to
mitigate this, with strong emission from the polar lobes in a few
transitions of H$_2$ and [Fe~{\sc ii}] (Smith \& Davidson 2001).
Furthermore, lower extinction in the IR allows us to peer {\it inside}
the Homunculus to study embedded emission structures.

Spatially resolved long-slit spectroscopy allows us to view the
reflected stellar spectrum from multiple directions.  Most previous
investigations of $\eta$ Car's IR spectrum have integrated light from
the entire Homunculus (Whitelock et al.\ 1983; Allen, Jones, \& Hyland
1985; McGregor, Hyland, \& Hillier 1988) or the bright unresolved core
(Hamann et al.\ 1994). Hamann et al.\ presented the optical to far-red
spectrum observed in the SE lobe, revealing a reflected broad-line
spectrum of the central star.  Long-slit spectra presented here
provide multiple viewing angles to $\eta$ Car for the first time at
wavelengths longer than 1 $\micron$.  $\eta$ Car was observed in March
2001, during its high-excitation state between ``spectroscopic
events'' that repeat every 5.5 years (Damineli 1996).  Latitudinal
structure of the wind changes dramatically during this cycle (Smith et
al.\ 2003), so these IR spectra also provide a benchmark for future
spectroscopy.

New observations obtained at CTIO are presented in \S 2.  Intrinsic IR
emission lines are used to derive the structure of the Homunculus in
\S 3, and in \S 4 and \S 5 other novel aspects of the IR spectra are
presented.  Finally, the central star, nearby ejecta, and reflected
spectra are discussed in \S 6.

\begin{figure*}\begin{center}
\epsfig{file=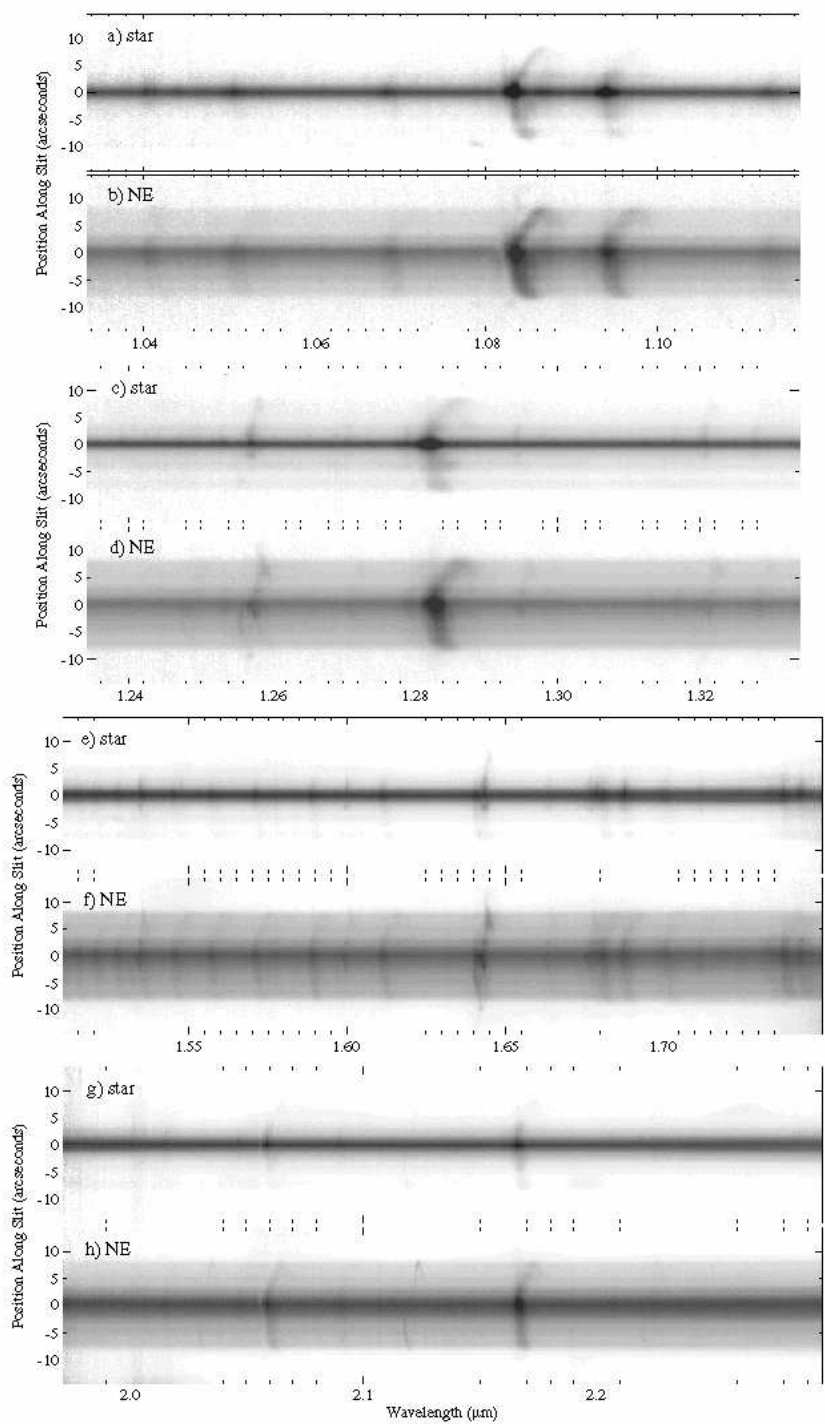,height=7.5in}\end{center}
\caption{Grayscale representations of long-slit OSIRIS spectra with the
slit centered on the star (at P.A.=-50$\arcdeg$), as well as one
offset position to the north-east (see Figure 1).  Panels (a) and (b)
show spectra in the I-band for the star and the NE position,
respectively.  Panels (c) and (d) show spectra in the J-band, panels
(e) and (f) show the H-band, and (g) and (h) show the K-band.}
\end{figure*}

\section{OBSERVATIONS}

Long-slit spectra of $\eta$ Carinae from 1 to 2.3 $\micron$ were
obtained on 2001 March 13 and 14 using
OSIRIS\footnotemark\footnotetext{See \tt
http://www.ctio.noao.edu/instruments/ir\_instruments
/osiris/index.html} mounted on the CTIO 4m telescope.  Data were
obtained in high-resolution mode, with an effective 2-pixel spectral
resolution of about 100 km s$^{-1}$ ($R \ \sim \ 3000$).  These
spectra are similar to observations obtained a year earlier in March
2000, for which preliminary results have already been presented (Smith
2001; Smith \& Davidson 2001).  However, the new data obtained in
March 2001 are superior for several reasons, including better seeing
($\sim$0$\farcs$4), and the alignment of the long slit aperture with
the polar axis of the Homunculus at P.A.=$-$50$\arcdeg$ (see Figure
1).  Spectra were obtained with the $\sim$0$\farcs$5-wide slit
aperture centered on the central star, as well as two positions offset
$\sim$1$\arcsec$ to the NE and SW.  A few spectra were also obtained
with the slit oriented perpendicular to the polar axis of the
Homunculus (labeled as `EQ1' and `EQ2' in Figure 1).  Sky-subtraction
was performed by chopping along the slit with offsets greater than the
size of the Homunculus.  Wavelengths were calibrated using
observations of an internal emission lamp.  Flux calibration and
telluric absorption correction were accomplished with reference to
spectroscopic standard stars.

Figure 2 shows long-slit spectra of the Homunculus in the I, J, H, and
K bandpasses observable from the ground, with long-slit spectra
centered on the star and 1$\arcsec$ NE (long-slit spectra for
positions 1$\arcsec$ SW are similar).  NE positions show fainter
emission features.  It is clear from Figure 2 that the observed
spectrum varies with position along the slit.  There are essentially
two different types of emission features present: 1) narrow emission
lines emitted by gas in the Homunculus, and 2) broad emission lines in
the wind of the central star reflected toward us by dust.  The
$\sim$0$\farcs$4 seeing for these observations reveals complex spatial
structure along the slit length, useful for future comparison with
{\it HST}/STIS data. Various details concerning the spatial variation
of these two types of features are discussed below.

\begin{figure}\begin{center}
\epsfig{file=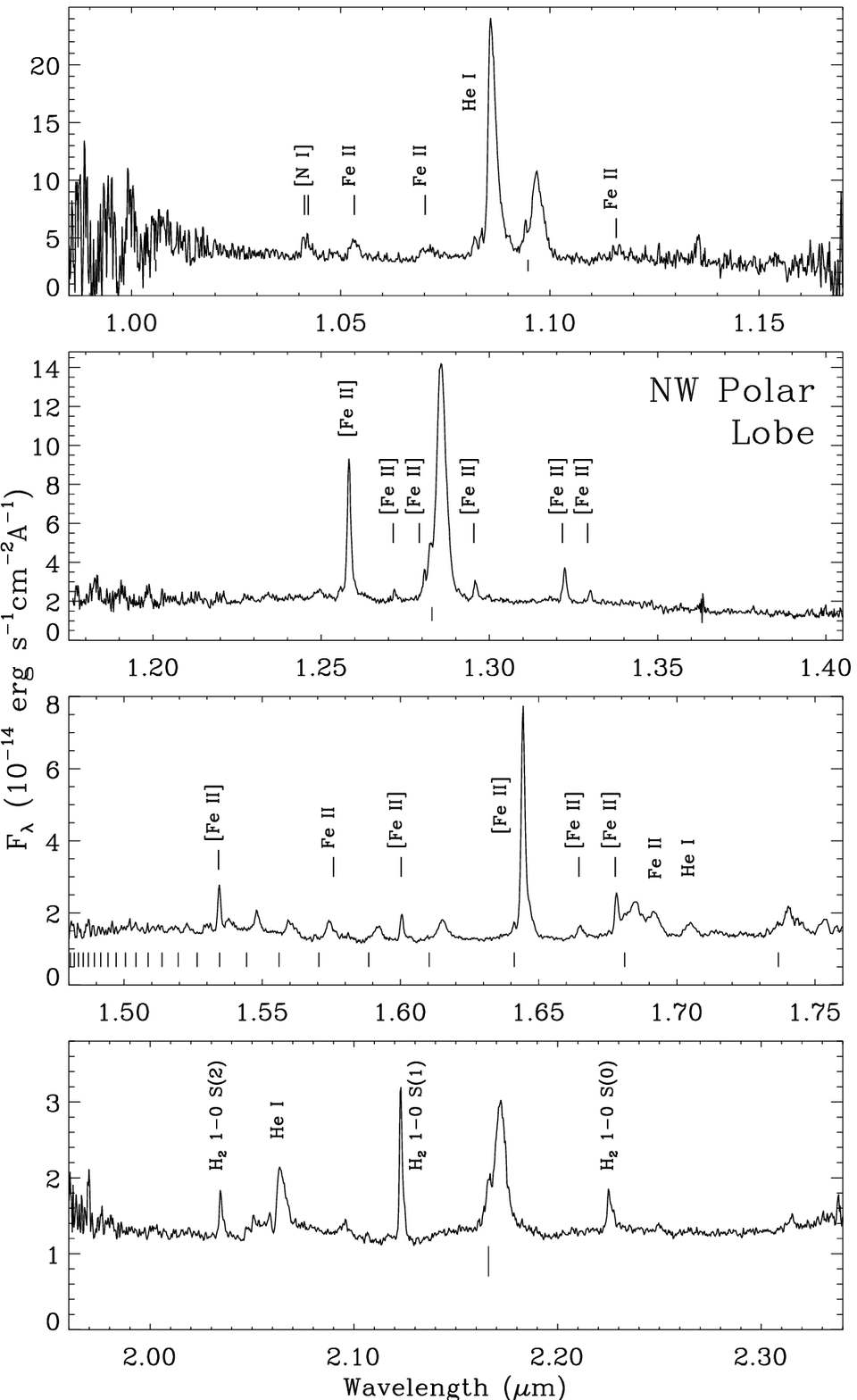,width=3.25in}\end{center}
\caption{Near-IR spectrum near the edge of the NW polar lobe of the
Homunculus, extracted from a 1$\farcs$2 segment of the 0$\farcs$5-wide
slit.  The extraction region is centered $\sim$7$\farcs$5 NW of the
star along the NE slit position (see Figure 1).  Rest wavelengths of
hydrogen lines are marked with dashes below the continuum, although
reflected stellar wind lines are redshifted.}
\end{figure}

\section{SHAPE AND STRUCTURE OF EMISSION FROM THE POLAR LOBES}

Proper motions for expanding debris in the Homunculus have been
measured by several authors (Gaviola 1950; Ringuelet 1958; Gehrz \&
Ney 1972; Currie et al.\ 1996; Smith \& Gehrz 1998; Morse et al.\
2001).  The age of $\eta$ Car's equatorial skirt is controversial, and
there is reason to suspect a mixture of material from multiple events
(Smith \& Gehrz 1998; Zethson et al.\ 1999; Morse et al.\ 2001;
Davidson et al.\ 2001).  The origin of the polar lobes is less
ambiguous --- various studies infer ejection within a few years of the
Great Eruption ($\sim$1843).  Since the polar lobes appear to be the
product of a single ejection event, apparent Doppler velocities can be
used to deduce their 3-D geometry.  Detailed results of previous
studies that used Doppler velocities of optical lines or polarization
measurements have shown some disagreement (Thackeray 1951, 1956$a$,
1956$b$; 1961; Meaburn et al.\ 1987, 1993; Hillier \& Allen 1992;
Allen \& Hillier 1993; Hillier 1997; Currie \& Dowling 1999;
Schulte-Ladbeck et al.\ 1999).  A recent study using {\it HST}/STIS
spectroscopy has improved the situation somewhat (Davidson et al.\
2001).  No attempt has yet been made to investigate this using IR
emission lines, where extinction is less problematic.  Recently, Smith
\& Davidson (2001) discovered bright, IR lines from H$_2$ and [Fe~{\sc
ii}] emitted by gas in the polar lobes.  These lines are obvious in
the spectrum near the edge of the NW lobe shown in Figure 3.
Molecular hydrogen lines in particular are ideal for investigating the
geometry, since they are {\it only} emitted by gas in the lobes and
therefore do not suffer from confusing velocity components reflected
by dust in the Homunculus, and because extinction at 2 $\micron$ is
low.  [Fe~{\sc ii}] lines also provide essential information.

\begin{figure*}\begin{center}
\epsfig{file=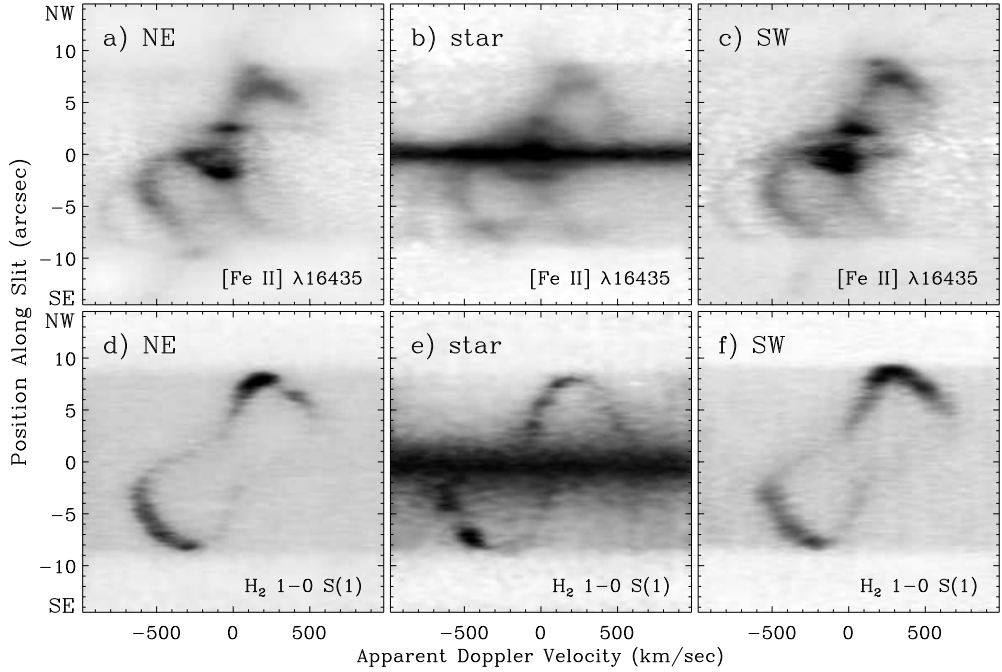,height=3.5in}\end{center}
\caption{Details of velocity structure for three slit positions in
Figure 1, for [Fe~{\sc ii}] $\lambda$16435 ($a$, $b$, $c$), and
H$_2$ 1$-$0 S(1) $\lambda$21218 ($d$, $e$, $f$). }
\end{figure*}

\subsection{Emission Structure}

Figure 4 shows velocity structure\footnotemark\footnotetext{Continuum
emission has been suppressed in Figure~4.  A smooth continuum model
was used for H$_2$, and a segment of the H-band spectrum around Br13
was used for [Fe~{\sc ii}] $\lambda$16435.} as a function of position
along the slit for [Fe~{\sc ii}] $\lambda$16435 and H$_2$ $v$=1-0 S(1)
$\lambda$21218.  Related lines, such as [Fe~{\sc ii}] $\lambda$12567
and H$_2$ $v$=1-0 S(2) $\lambda$20338 (see Figures 2 and 3), show
similar structure but are fainter.  Three slit positions are shown,
with the slit crossing through the star and offset positions NE and
SW.  Short exposure times had to be used with the slit centered on the
star, so the signal-to-noise there is worse, but extended velocity
structure can still be distinguished.

[Fe~{\sc ii}] emission components in Figure 4 are complex and trace
several different features in the Homunculus, but can be understood in
the context of previous work (see Allen \& Hillier 1993; Davidson et
al.\ 2001; Ishibashi et al.\ 2003).  By comparison, H$_2$ emission
structure is surprisingly elegant, and seems to trace only material in
the polar lobes.  Lower extinction at IR wavelengths has allowed us to
{\it see the back wall of the SE lobe} unambiguously for the first
time in both H$_2$ and [Fe~{\sc ii}].  Also, subtle differences in
shape can be seen from one position to the next, especially in H$_2$,
and there is a conspicuous gap in the emission structure at 6$\arcsec$
SE for the slit position passing through the star (Figures 4$b$ and
$e$).  This slit also traverses an infamous ``hole'' in the SE lobe
seen in images of the Homunculus (see Figure 1), the nature of which
has been controversial (Smith et al.\ 1998; Morse et al.\ 1998;
Schulte-Ladbeck et al.\ 1999).  The fact that both [Fe~{\sc ii}] and
H$_2$ emission show a pronounced gap in the emission structure,
combined with the fact that we apparently see the back wall of the
Homunculus at the same position, has interesting implications.

The images in Figure 4 are stretched horizontally from their original
pixel sampling to reflect the actual shape of the polar lobes, but the
magnification factor is not arbitrary; it depends on assumed values
for the heliocentric distance and ejecta age.  Defining $Z$ as the
position along the line-of-sight in arcseconds, apparent Doppler
velocities $v$ in km s$^{-1}$ yield values for $Z$ if the age of the
ejecta $t_{yr}$ and heliocentric distance $D_{pc}$ are known using the
relation

\begin{equation}
Z  =   0.21 \frac{ v \ t_{yr} } {D_{pc}} \ arcseconds.
\end{equation}

\noindent Figures 5$a$ and $b$ show how emission structure varies with
$Z$ and position along the slit for a heliocentric distance of 2250 pc
and an age of 158 years (ejection date of 1843), using the same length
scale for both the vertical and horizontal axes (the same proportions
were used in Figure 4 as well, although Doppler velocities were shown
there).  A distance of 2250 pc yielded the most symmetric appearance
for the polar lobe structure in the H$_2$ line; assumed values of 2200
and 2300 pc were noticably distorted.  Davidson et al.\ (2001) found
the same result for optical emission lines in {\it HST}/STIS data, and
their quoted uncertainy of $\pm$50 pc is applicable here as
well. Figures 5$a$ and $b$ show a cross section of the Homunculus in
[Fe~{\sc ii}] $\lambda$16435 and H$_2$ $\lambda$21218 emission for an
average of the NE and SW slit positions.

Figures 5$c$ and $d$ compare the relative distribution of [Fe~{\sc
ii}] and H$_2$ by superimposing isophotal contours of one line's
velocity structure over grayscale images of the other.  It is clear
from these figures that [Fe~{\sc ii}] emission associated with the
polar lobes resides mostly interior to the H$_2$.  Figure 6 conveys
the same conclusion with tracings of the velocity profile for each
line at a few representative positions. The particular spatial
distribution of these two lines, combined with the fact that neither
emission line appears to exceed the confines of the polar lobes,
supports an earlier contention (Smith \& Davidson 2001) that this
emission may arise from a shock caused by a fast stellar wind
impacting the {\it inside} surfaces of the polar
lobes.\footnotemark\footnotetext{[Fe~{\sc ii}] emission gives evidence
for a shock at the {\it outside} surface of the polar lobe as well;
there is a small emission blob at the edge of the NW lobe, seen most
clearly in Figure 5$a$.  This blob is clearly separated from [Fe~{\sc
ii}] inside the NW lobe, and the gap between them corresponds
precisely to the location of the H$_2$ emission (Figure 5$c$).  The
spectrum of this outer shock and its interesting implications will be
discussed in a later publication.}

\begin{figure*}\begin{center}
\epsfig{file=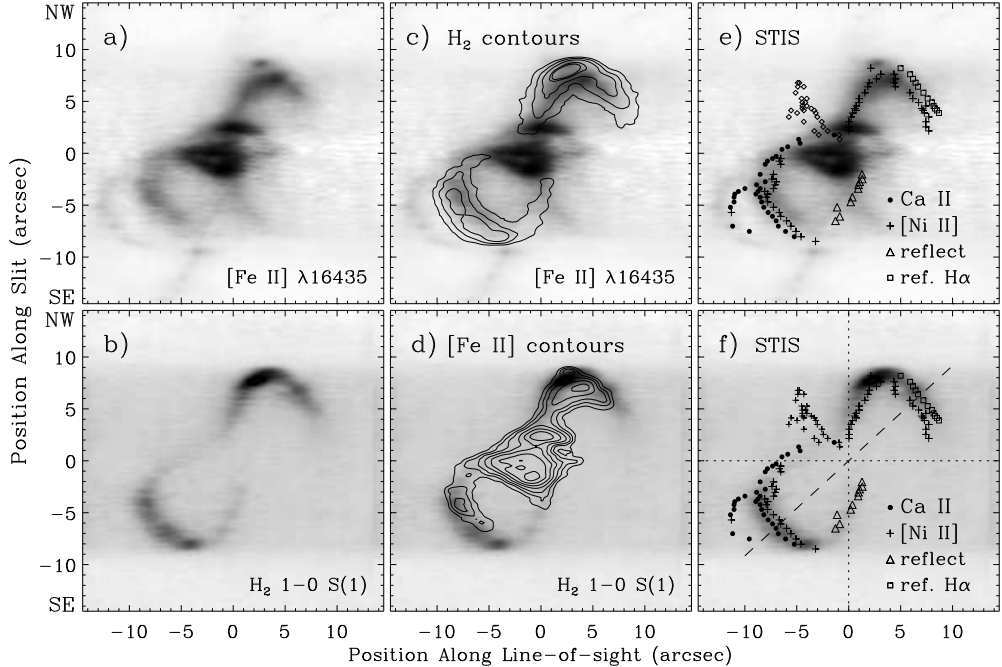,height=3.5in}\end{center}
\caption{Panels (a) and (b) show emission-line structure for
[Fe~{\sc ii}] $\lambda$16435 and H$_2$ 1$-$0 S(1) 2.1218 $\micron$,
respectively, with the horizontal axis scale determined using equation
(1).  The observed structure represents a meridional cross section of
the Homunculus for an average of the NE and SW slit positions from
Figure 4.  Panels (c) and (d) are the same as (a) and (b), but with
isophotal contours of the complementary emission line superimposed.
Panels (e) and (f) are the same as (a) and (b), overplotted with
points derived from velocities measured in {\it HST}/STIS data by the
author (see also Davidson et al.\ 2001).  The derived structure is
plotted for Ca~{\sc ii} $\lambda\lambda$3935,3970 absorption, [Ni~{\sc
ii}] $\lambda$7379 emission, and reflected emission from [Ni~{\sc ii}]
and H$\alpha$ corrected for expansion of reflecting dust.}
\end{figure*}

\begin{figure}\begin{center}
\epsfig{file=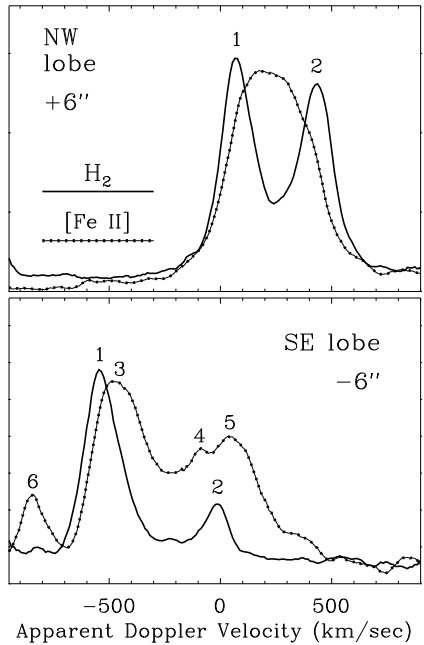,height=3.4in}\end{center}
\caption{Tracings of H$_2$ $\lambda$21218 and [Fe~{\sc ii}]
$\lambda$16435 profiles for a 0$\farcs$5 segment along the slit at
positions in the NW polar lobe (top) and the SE polar lobe (bottom).
Components 1 and 2 represent H$_2$ emission from the front and back of
the lobes, respectively.  [Fe~{\sc ii}] from the front and back of the
NW lobe is unresolved, but it is in the SE lobe (components 3 and 4).
Component 5 is reflected emission, and component 6 corresponds to
high-speed gas outside the Homunculus (see Figure 4).}
\end{figure}

Doppler velocities for optical lines in {\it HST}/STIS data (measured
independently by the author, but see Davidson et al.\ 2001) have been
converted to the same coordinates along the line-of-sight and plotted
over the IR emission structure in Figure 5$e$ and $f$.  Ca~{\sc ii}
absorption in STIS data appears to overlap with infrared H$_2$
emission in the SE lobe in Figure 5$f$ (some Ca~{\sc ii} absorption
also arises at high velocities in front of the SE lobe, associated
with the [Fe~{\sc ii}] emission features in ejecta outside the
Homunculus).  [Ni~{\sc ii}] $\lambda$7379 emission overlaps with
[Fe~{\sc ii}] $\lambda$16435 instead, and resides interior to both
H$_2$ emission and Ca~{\sc ii} absorption.  Reflected emission lines
from the back walls of the polar lobes trace dust, which seems to
reside in the exterior parts of the lobes coincident with H$_2$.
Therefore, H$_2$ emission appears to be the best unambiguous tracer of
the shape and orientation of the dense neutral gas and dust, while
atomic emission lines arise from material inside the lobes.  This is
why Davidson et al.'s ``model 1'' derived from [Ni~{\sc ii}]
$\lambda$7379 produced a shape that was too small compared to
continuum images.

\begin{table}\begin{center}
\caption{R($\theta$) and V($\theta$) for the Polar Lobes}
\begin{tabular}{@{}lcc} \hline\hline
Latitude &Radius &Velocity \\ (deg) &(AU) &(km s$^{-1}$) \\ \hline

5		&4890	&148		\\
10		&5220	&158		\\	
15		&5540	&168		\\
20		&5980	&181		\\
25		&6520	&197		\\
30		&7120	&215		\\
35		&8150	&246		\\
40		&9670	&292		\\
45		&11950	&361		\\
50		&16030	&485		\\
55		&17930	&542		\\
60		&18800	&568		\\
65		&19670	&595		\\
70		&20650	&625		\\
75		&21350	&646		\\
80		&21140	&639		\\
85		&21030	&636		\\
90		&21080	&637		\\	\hline

\end{tabular}
\end{center}
\end{table}

\subsection{Geometry of the Homunculus}

Various models for the exact 3-D geometry of the Homunculus have been
controversial, since the detailed shape of the polar lobes may hold
important clues to ejection physics during the Great Eruption.  The
shape of the lobes depends on the tilt angle of the polar axis with
respect to our line-of-sight; to avoid confusion, the inclination
angle $i$ is used as defined for binary systems, where a view from the
equator is $i$=90$\arcdeg$.  The inclination was measured by rotating
Figure 5$b$ so that the polar axis was nearly vertical, and then
flipping the image about the horizontal and vertical axes and examing
the lobe thickness when the images were added.  The smallest
dispersion (i.e. the most symmetric shape) was found for
$i$=42$\fdg$5$\pm$2$\arcdeg$.  This value is only slightly larger than
$i\approx$41$\arcdeg$ found by Davidson et al.\ (2001), mainly because
of differences in the shape of H$_2$ emission compared with optical
lines.  Smith et al.\ (1999) also derived $i\approx$40$\arcdeg$ from
equatorial features seen in IR images.  Figure 7 shows the end result
of rotating Figure 5$b$ by 90$-i$, reflecting it across the various
symmetry axes, and averaging.  This gives an artificially symmetric
picture of the polar lobes, but it is a good approximation for
discussion here.  The size of the nebula is derived assuming a
distance of 2250 pc and an age of 158 years.  Table 1 lists
representative values for the radius and expansion velocity at each
latitude ($\theta$=90$\arcdeg$ at the pole).

The Homunculus shape is often discussed in the context of three
geometric models, which approximate the lobes as a pair of flasks,
bubbles (spheres), or bipolar caps (see Hillier 1997).  None of these
three models adequately describes the geometry in Figure 7, but one
can see how each has a kernal of truth.  One important way that Figure
7 differs from all these models is that near the equator, the walls of
the polar lobes clearly do not converge on the central star.  Instead,
the two lobes appear meet a few thousand AU from the star. One might
expect an equatorial torus to reside at the intersection of these
lobes, and in fact, a structure resembling a disrupted torus is seen
there in thermal-IR images (Smith et al.\ 2002).

Figure 7 seems generally consistent with hydrodynamic simulations of
bipolar nebula formation applied to $\eta$ Car (e.g. Frank et al.\
1995, 1998; Dwarkadas \& Balick 1998), and some details are as well.
For instance, the scenario discussed by Frank et al.\ (1995) predicts
two shocks -- one at the inside of the polar lobes as the
post-eruption wind collides with ejecta, and one on the outside as
ejecta sweep up the ambient medium.  [Fe~{\sc ii}] $\lambda$16435
shows emission from both inner and outer shocks in Figure 5, as
mentioned earlier.

\begin{figure}\begin{center}
\epsfig{file=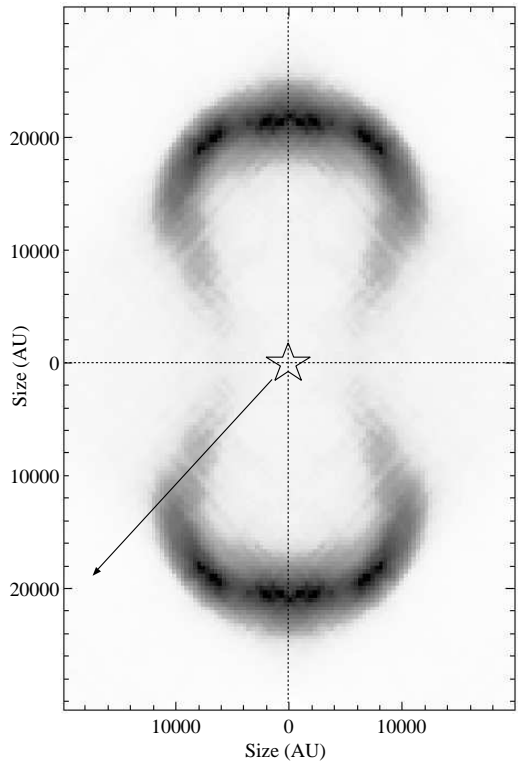,height=4in}\end{center}
\caption{Meridional cross-section of the shape of the polar lobes in
the Homunculus, derived by rotating the H$_2$ emission structure in
Figure 5$b$ counter-clockwise by 90$-i$, and reflecting it about the
symmetry axes.  Measured parameters as functions of latitude are given
in Table 1.  The arrow shows the direction toward the observer.  The
lobes are $\sim$5\% smaller for a figure prepared the same way using
the [Fe~{\sc ii}] $\lambda$16435 line (not shown).}
\end{figure}

\begin{figure}\begin{center}
\epsfig{file=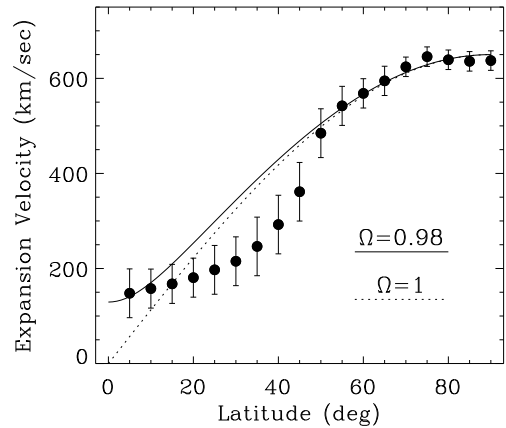,width=3in}\end{center}
\caption{Expansion velocity of H$_2$ in the polar lobes as a function
of latitude (see Table 1).  Solid and dotted curves show equation
(2) for two values of $\Omega$ and a polar velocity of 650 km
s$^{-1}$.}
\end{figure}

The observed structure does not necessarily rule out other ideas for
the formation of the Homunculus.  After examining the geometry of the
present-day aspherical stellar wind, Smith et al.\ (2003) suggest
that the shape of the Homunculus might have arisen from intrinsically
asymmetric ejection during the Great Eruption, and may have been
shaped somewhat by an asymmetric post-eruption wind.  In that case, we
might expect the latitudinal dependence of ejection velocity during
the eruption to be related to the star's rotation.  For instance,
ejection speed $v_{\infty}$ at various latitudes $\theta$ may vary
with escape speed on a rotating star as

\begin{equation}
v_{\infty} = v_{pole} \big{(} 1 - \Omega^2 \cos^2 \theta \big{)}^{\frac{1}{2}}
\end{equation}

\noindent where $\Omega \equiv (v_{rot} / v_{esc})$ at the
equator.  Expansion speeds at each latitude in Table 1 are plotted in
Figure 8, with curves of equation (2) for values of $\Omega$ near
critical rotation.  Equation (2) provides an adequate description of
the shape of the polar lobes at latitudes above $\theta$=50$\arcdeg$,
where most of the mass was presumably ejected in thick polar caps
(e.g., Hillier \& Allen 1992).  The walls of the polar lobes at
$\theta < 50\arcdeg$ are optically thinner and presumably less
massive; they may have been decelerated as they plowed through the
ambient medium, or during a hypothetical collision with a pre-existing
torus (e.g., Dwarkadas \& Balick 1998).

Figures 4 and 5 also show narrow [Fe~{\sc ii}] emission from a fast
outer bubble projected in front of the SE polar lobe.  This outer
bubble is seen at $-$500 to $-$900 km s$^{-1}$ in [Fe~{\sc ii}], and
appears at slightly lower velocities in Ca~{\sc ii} {\it absorption}
in STIS spectra (Figures 5$e$ and $f$).  Apparently, it also causes
absorption in the He~{\sc i} $\lambda$10830 line (see \S 6.3 and
Figure 20 below).  This outer feature corresponds to velocity
component number 6 in tracings of [Fe~{\sc ii}] in Figure 6.  Its
actual 3-D structure is uncertain because its age is not known.

\begin{table*}
\begin{minipage}{6in}
\caption{Intrinsic Emission-lines in the Homunculus$^a$}
\begin{tabular}{@{}llccc} \hline\hline
Wavelength	&I.D.$^{b}$				&NW Lobe Flux		&LH Flux		&Fan Flux	\\	
(\AA)		&	&(erg s$^{-1}$ cm$^{-2}$)	&(erg s$^{-1}$ cm$^{-2}$)	&(erg s$^{-1}$ cm$^{-2}$) \\ \hline

12567		&[Fe~{\sc ii}] $(a^6D - a^4D)$		&7.22$\times$10$^{-13}$	&5.80$\times$10$^{-12}$	&1.94$\times$10$^{-11}$	\\
12788		&[Fe~{\sc ii}] $(a^6D - a^4D)$		&4.83$\times$10$^{-14}$	&7.92$\times$10$^{-13}$	&3.28$\times$10$^{-12}$	\\
12818		&H~{\sc i} Pa$\beta$		&($<$4$\times$10$^{-15}$) &($<$2$\times$10$^{-14}$)	&2.69$\times$10$^{-11}$	\\
12943		&[Fe~{\sc ii}] $(a^6D - a^4D)$		&8.84$\times$10$^{-14}$	&9.65$\times$10$^{-13}$	&3.06$\times$10$^{-12}$	\\
13205		&[Fe~{\sc ii}] $(a^6D - a^4D)$		&2.23$\times$10$^{-13}$	&1.13$\times$10$^{-12}$	&3.91$\times$10$^{-12}$	\\
15335		&[Fe~{\sc ii}] $(a^4F - a^4D)$		&1.57$\times$10$^{-13}$	&1.27$\times$10$^{-12}$	&6.96$\times$10$^{-12}$	\\
15995		&[Fe~{\sc ii}] $(a^4F - a^4D)$		&9.92$\times$10$^{-14}$	&1.33$\times$10$^{-12}$	&5.71$\times$10$^{-12}$	\\
16435		&[Fe~{\sc ii}] $(a^4F - a^4D)$		&1.05$\times$10$^{-12}$	&7.92$\times$10$^{-12}$	&1.89$\times$10$^{-11}$	\\
16637		&[Fe~{\sc ii}] $(a^4F - a^4D)$		&7.44$\times$10$^{-14}$	&5.43$\times$10$^{-13}$	&2.64$\times$10$^{-12}$	\\
16769		&[Fe~{\sc ii}] $(a^4F - a^4D)$		&1.39$\times$10$^{-13}$	&1.18$\times$10$^{-12}$	&7.83$\times$10$^{-12}$	\\
16787		&Fe~{\sc ii} $(z^4F - c^4F)$		&...			&1.25$\times$10$^{-13}$	&6.92$\times$10$^{-12}$	\\
16873		&Fe~{\sc ii} $(z^4F - c^4F)$		&...			&2.68$\times$10$^{-13}$	&9.25$\times$10$^{-12}$	\\
17111		&[Fe~{\sc ii}] $(a^4F - a^4D)$		&...			&7.92$\times$10$^{-13}$	&1.26$\times$10$^{-12}$	\\
17414		&Fe~{\sc ii} $(z^4F - c^4F)$		&...			&4.56$\times$10$^{-13}$	&7.67$\times$10$^{-12}$	\\
19572		&Fe~{\sc ii} $(z^4F - c^4F)$		&...			&...			&6.61$\times$10$^{-12}$	\\
19746		&Fe~{\sc ii} $(z^4F - c^4F)$		&...			&2.90$\times$10$^{-13}$	&2.31$\times$10$^{-12}$	\\
20151		&[Fe~{\sc ii}] $(a^2G - a^2H)$		&...			&5.33$\times$10$^{-13}$	&1.94$\times$10$^{-12}$	\\
20338		&H$_2$ 1-0 S(2)				&1.36$\times$10$^{-13}$	&...			&...			\\
20460		&[Fe~{\sc ii}] $(a^4P - a^2P)$		&...			&6.08$\times$10$^{-13}$	&1.50$\times$10$^{-12}$	\\
20581,600	&He~{\sc i}, Fe~{\sc ii} $(z^4F - c^4F)$ &...			&3.74$\times$10$^{-13}$	&3.61$\times$10$^{-12}$	\\
20888		&Fe~{\sc ii} $(z^4F - c^4F)$		&2.20$\times$10$^{-14}$	&3.49$\times$10$^{-13}$	&2.96$\times$10$^{-12}$	\\
21218		&H$_2$ 1-0 S(1)				&4.17$\times$10$^{-13}$	&...			&...			\\
21655		&H~{\sc i} Br$\gamma$		&($<$4$\times$10$^{-15}$)    &($<$3$\times$10$^{-14}$)	&5.49$\times$10$^{-12}$	\\
22233		&H$_2$ 1-0 S(0)				&1.51$\times$10$^{-13}$	&...			&...			\\
22477		&H$_2$ 2-1 S(1)				&1.77$\times$10$^{-14}$	&...			&...		\\ \hline

\end{tabular}

$^a$See Figures 3, 10, and 11. If more than one wavelength is given,
the line is blended, and both lines are expected to contribute
significantly.

$^b$J subscripts for Fe$^+$ lines have been omitted.
\end{minipage}
\end{table*}

\subsection{Excitation of the Polar Lobes}

Observed fluxes from Figure 3 for several {\it intrinsic} emission
lines in the NW polar lobe are listed in Table
2.\footnotemark\footnotetext{Since the spectrum of the NW polar lobe
is a complex mix of intrinsic emission and reflected emission lines
from the stellar wind and Weigelt blobs (redshifted by almost $+$1000
km s$^{-1}$), some lines in the spectrum are excluded and corrections
were applied after considering kinematic information.  For instance,
[Fe~{\sc ii}] $\lambda$16435 is blended with Br12, and the flux for
this line was adjusted accordingly in Table 2.  Also, Br$\gamma$ has
an intrinsic narrow component in addition to the strong reflected line
from the stellar wind, but this narrow component is blueshifted and
appears to be emitted by slow-moving equatorial gas (see \S 5).
Therefore, Table 2 lists a likely upper limit to the Br$\gamma$ flux
in the polar lobe.}  H$_2$ $v$=2-1 and Q-branch transitions are absent
or very weak in Figure 3; these are typically weak in molecular gas
excited by shocks (e.g., Shull \& Hollenbach 1978), and relatively
strong H$_2$ 1-0 S(1) emission (low excitation temperature) is
conventionally taken to indicate shock excitation.  Here, the H$_2$
1-0 S(1) / H$_2$ 2-1 S(1) ratio of $\sim$12 implies a characteristic
excitation temperature of $\sim$2000 K.  This rules out {\it
radiative} fluorescence (i.e. direct excitation through absorption in
the Lyman and Werner bands in the UV) as the excitation mechanism for
the H$_2$, since that mechanism results in a H$_2$ 1-0 S(1) / H$_2$
2-1 S(1) ratio of $\sim$2 (Sternberg \& Dalgarno 1989).  However,
Sternberg \& Dalgarno (1989) point out that {\it collisional}
fluorescent emission can also result in low excitation temperatures
($\sim$10$^3$ K) in dense molecular gas irradiated by a strong UV
flux.  Sternberg \& Dalgarno stress that collisional excitation of
dense gas resulting from radiative heating can mimic line ratios in
shocks.

A similar conundrum may exist for the excitation mechanism of [Fe~{\sc
ii}] lines at 12567 and 16435 \AA.  Models for IR emission from dense
interstellar shocks (e.g., McKee, Chernoff, \& Hollenbach 1984)
predict enhanced [Fe~{\sc ii}] compared to H.  Observationally,
shock-excited gas such as that in supernova remnants shows high ratios
of [Fe~{\sc ii}] $\lambda$16435/Br$\gamma$ $\gtrsim$30 (Graham et al.\
1987; Oliva et al.\ 1990; Seward et al.\ 1983), whereas H~{\sc ii}
regions have [Fe~{\sc ii}] $\lambda$16435/Br$\gamma$ $\la$1 (Moorwood
\& Oliva 1988; Mouri et al.\ 1990).  The observed [Fe~{\sc ii}]
$\lambda$16435/Br$\gamma$ ratio in the NW polar lobe is $>$26 (Table
3), which would conventionally be interpreted as a shock. However,
radiation may be able to heat the gas to a few 10$^3$ K as well,
allowing collisionally-excited [Fe~{\sc ii}] lines and still
suppressing the H recombination lines.  For instance, the Crab Nebula
has a high [Fe~{\sc ii}] $\lambda$16435/Br$\gamma$ ratio ranging from
10 to 50 (Graham et al.1990), and it is photoexcited by a power-law
synchrotron spectrum.  The excitation of this warm gas in the polar
lobes of $\eta$ Car will be investigated quantitatively by Ferland,
Davidson, \& Smith (2002).

In any case, it seems clear from Figure 5 that [Fe~{\sc ii}] emission
arises at the {\it inside} surface of the hollow polar lobes; peaks in
H$_2$ and [Fe~{\sc ii}] are separated by $\sim$1000 AU.  The
separation of H$_2$ and [Fe~{\sc ii}] emission may have implications
for grain destruction and the gas-phase iron abundance in this
photodissociation region.

[Fe~{\sc ii}] $\lambda$12567 and $\lambda$16435 share the same upper
level, and atomic physics predicts an intrinsic flux ratio of
$\lambda$12567/$\lambda$16435 \ $\approx$ \ 1.36 (Nussbaumer \& Storey
1988).  The observed flux ratio less than 1 in Table 3 therefore
indicates considerable reddening from circumstellar dust in the walls
of the polar lobes; however, the reddening law for dust in the
Homunculus is anomolous and poorly constrained, so it is not possible
to extrapolate this ratio to $A_V$.

If $\eta$ Car lost $\sim$2 $M_{\odot}$ during the Great Eruption
(Smith et al.\ 1998), then the particle density in the polar lobes
today should be on the order of 10$^{4.5}$ to 10$^5$ cm$^{-3}$, given
the apparent thickness and radii of the lobes in IR images (Smith et
al.\ 1998; 2002).  Therefore, the density is within the regime for
collisional fluorescence of H$_2$ discussed by Sternberg \& Dalgarno
(1989).  The [Fe~{\sc ii}]-emitting region seems to be spatially
segregated from H$_2$, and the electron density there can be estimated
from ratios of some infrared [Fe~{\sc ii}] lines with weak temperature
dependence that are all collisionally excited $a^4F-a^4D$ transitions
with closely spaced energy levels near the ground state.  Following
Nussbaumer \& Storey (1980; 1988), Table 3 lists values for $n_e$
derived from the diagnostic line ratios [Fe~{\sc ii}]
$\lambda$15335/$\lambda$16435, $\lambda$15995/$\lambda$16435, and
$\lambda$16637/$\lambda$16435.  The average of these is
$\sim$10$^{4.1}$ cm$^{-3}$ in the polar lobes, close to the expected
value mentioned above.  This implies that [Fe~{\sc ii}] in the lobes
arises in ejecta from the Great Eruption, rather than post-eruption
wind.

\begin{table}
\caption{Nebular Densities and Line Ratios}
\begin{tabular}{lcccc} \hline\hline
Diagnostic				&NW Lobe	&LH	&Fan	&Weigelt blobs	\\ \hline

log $n_e$ ($\frac{15335}{16435}$)$^{a}$	&4.0	&4.0	&$\gtrsim$5.3	&$>$5.8	\\
log $n_e$ ($\frac{15995}{16435}$)$^{a}$	&3.9	&4.4	&$\gtrsim$5.4	&$>$5.6	\\
log $n_e$ ($\frac{16637}{16435}$)$^{a}$	&4.2	&4.2	&$\gtrsim$5.0	&$>$5.8	\\
log $n_e$ (average)			&4.1	&4.2	&$\gtrsim$5.3	&$>$5.6	\\
$\lambda$16435/Br$\gamma$		&$>$26	&$>$26	&3.4		&0.74	\\
$\lambda$12567/$\lambda$16435		&0.7	&0.7	&1.0		&0.99	\\
\hline

\end{tabular}
$^{a}$See Nussbaumer \& Storey (1980, 1988).
\end{table}

\section{THE LITTLE HOMUNCULUS}

Ishibashi et al.\ (2003) discovered a bipolar nebula expanding inside
the Homunculus, based on spatially-resolved {\it HST}/STIS
spectroscopy.  Proper motions of structures in those data indicate
that this `Little Homunculus' seems to have an age of $\sim$100 years,
perhaps ejected during the minor eruption of $\eta$ Carinae in 1890.
The Little Homunculus (LH) can also be seen in some IR emission lines
like [Fe~{\sc ii}] $\lambda$12567 and $\lambda$16435 (Figure 9).  The
bipolar shape of the LH in [Fe~{\sc ii}] $\lambda$16435 is more
prominent than in optical spectra because collisionally-excited IR
lines like [Fe~{\sc ii}] $\lambda$16435 are so bright, and because
extinction from the SE lobe is less severe in the IR.  The hollow
structure of the LH's blueshifted SE lobe is clear, but the NW polar
lobe of the LH is fainter.  Equatorial dust seen in thermal-IR images
(Smith et al.\ 2002) may block light from the far side of the LH.
Polar ejecta in the LH expand at speeds around 200 km s$^{-1}$ in
Figure 9.  Walborn \& Liller (1977) pointed out that during the 1890
eruption, absorption features were seen at $-$200 km s$^{-1}$ in
$\eta$ Car's spectrum (Whitney 1952).  This supports Ishibashi et
al.'s claim that the LH may have originated in that event.

Figure 10 shows the IR spectrum at the brightest position in the LH,
indicated by the arrow in Figure 9.  [Fe~{\sc ii}] lines are enhanced,
and most show a blueshifted component at about $-$500 km s$^{-1}$ from
the SE lobe of the Homunculus.  H lines in Figure 10 are reflected
from the wind and Weigelt blobs, and are not relevant to the LH
itself.  Table 2 lists intensities for emission arising in the LH;
emission components from the Homunculus (including H$_2$) and
reflected wind lines are excluded.

\begin{figure}\begin{center}
\epsfig{file=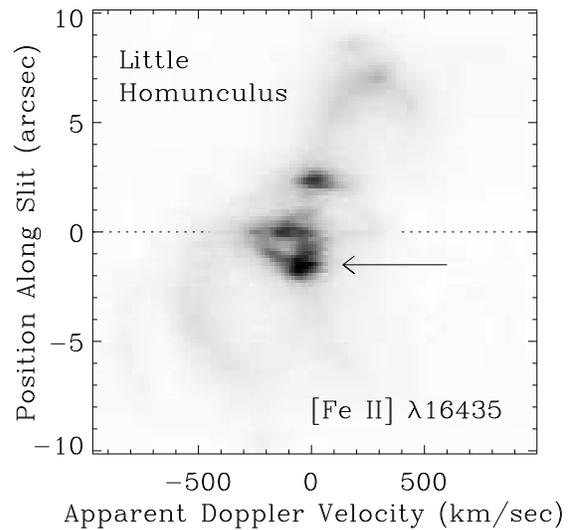,width=2.8in}\end{center}
\caption{Same as Figure 5$a$, with the display chosen to highlight
bright [Fe~{\sc ii}] $\lambda$16435 emission from the `Little
Homunculus'.  Apparent Doppler velocities are shown, but the
horizontal scale is stretched by a factor corresponding to an ejection
date of 1890 (see equation 1).  The arrow shows the location of the
extracted spectrum shown in Figure 10.}
\end{figure}

\begin{figure}\begin{center}
\epsfig{file=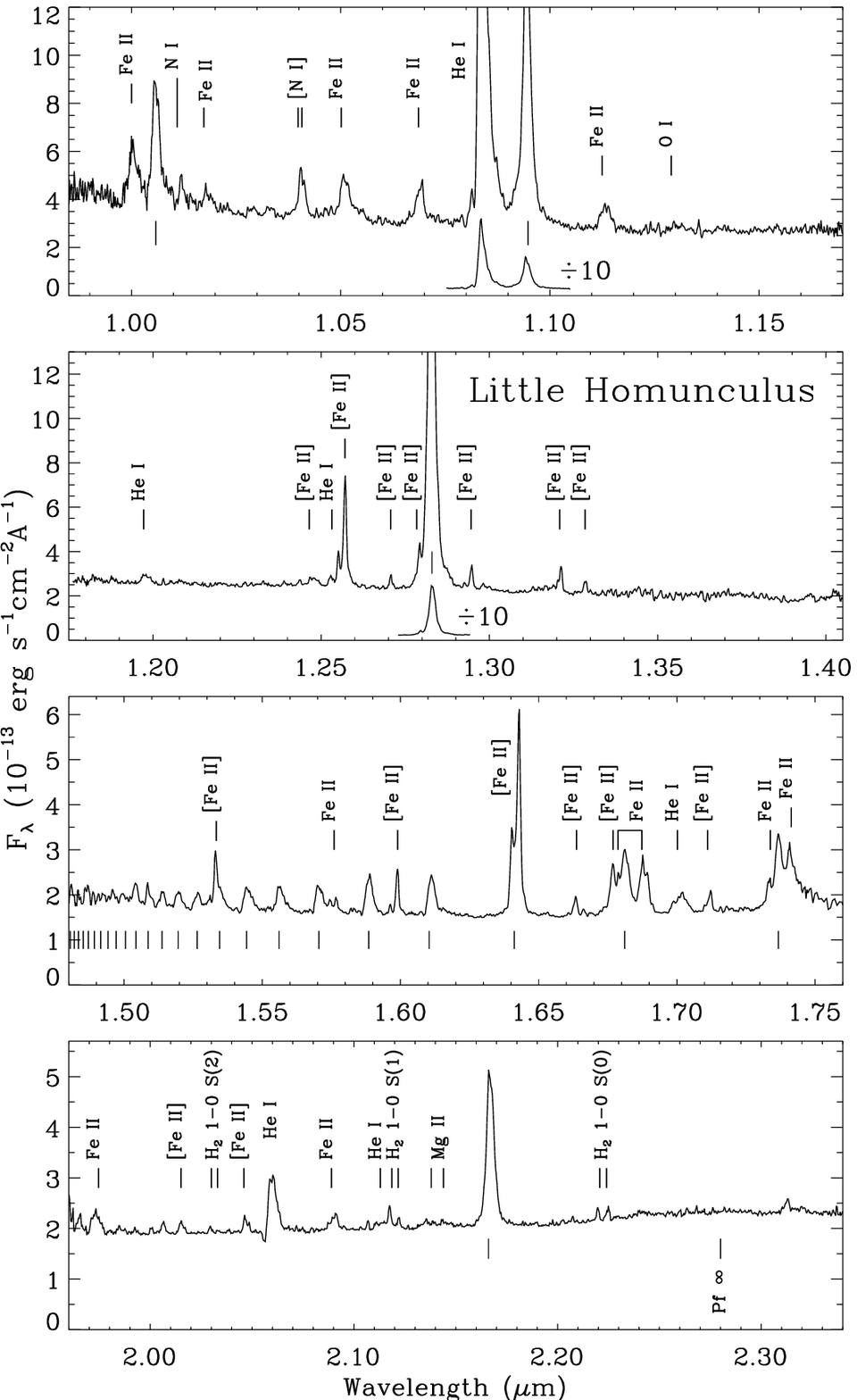,width=3.25in}\end{center}
\caption{Near-IR spectrum of the brightest part of the Little
Homunculus, extracted from a 1$\arcsec$ segment of the 0$\farcs$5-wide
slit aperture, centered 1$\farcs$5 SE of the minor axis of the
Homunculus at the position of the arrow in Figure 9.  The spectrum was
taken with the slit at the NE offset position (see Figure 1).
Hydrogen lines are marked with a dash below the continuum level.}
\end{figure}

The LH spectrum is qualitatively similar to the spectrum of the NW
polar lobe, except that H$_2$ emission is absent here.  Most infrared
[Fe~{\sc ii}] lines in the LH spectrum arise from one of two
multiplets ($a^4F-a^4D$ or $a^6D-a^4D$), which are
collisionally-excited transitions with an upper level 1 eV from the
ground state.  The [Fe~{\sc ii}] $\lambda$16435/Br$\gamma$ ratio is
quite high, like that seen in the NW polar lobe (see Table 3), and
comments about the excitation mechanism described there may apply.
However, an important difference is that the LH seems to lack bright
thermal-IR emission from dust (Smith et al.\ 2002), as well as H$_2$.
A contribution from photoexcitation in the LH cannot be ruled out, but
the spectrum of the LH definitely lacks strong intrinsic IR emission
arising from Fe~{\sc ii} transitions pumped by UV fluorescence (see \S
5.1), which are prominent in some other ejecta near $\eta$ Car.

The [Fe~{\sc ii}] $\lambda$12567/$\lambda$16435 ratio in the LH is
nearly the same as for the NW lobe (Table 3), and so both suffer
similar reddening from dust in the lobes.  As with the polar lobe
spectrum, ratios of certain [Fe~{\sc ii}] lines from the $a^4F-a^4D$
multiplet are useful for estimating the electron density in the LH,
which appears to be $\sim$10$^{4.2}$ cm$^{-3}$ (Table 3).

\section{PECULIAR EQUATORIAL EMISSION}

It is interesting that both the LH and the polar lobes of the larger
Homunculus show strong intrinsic emission in collisionally-excited
lines.  These same tracers of collisional excitation are not seen in
emission from equatorial gas, which would be projected in front of the
NW lobe in Figures 4, 5, and 9. However, some lines do show
interesting evidence for blueshifted equatorial emission, providing
insight to the nature of $\eta$ Car's equatorial ejecta.

\begin{figure}\begin{center}
\epsfig{file=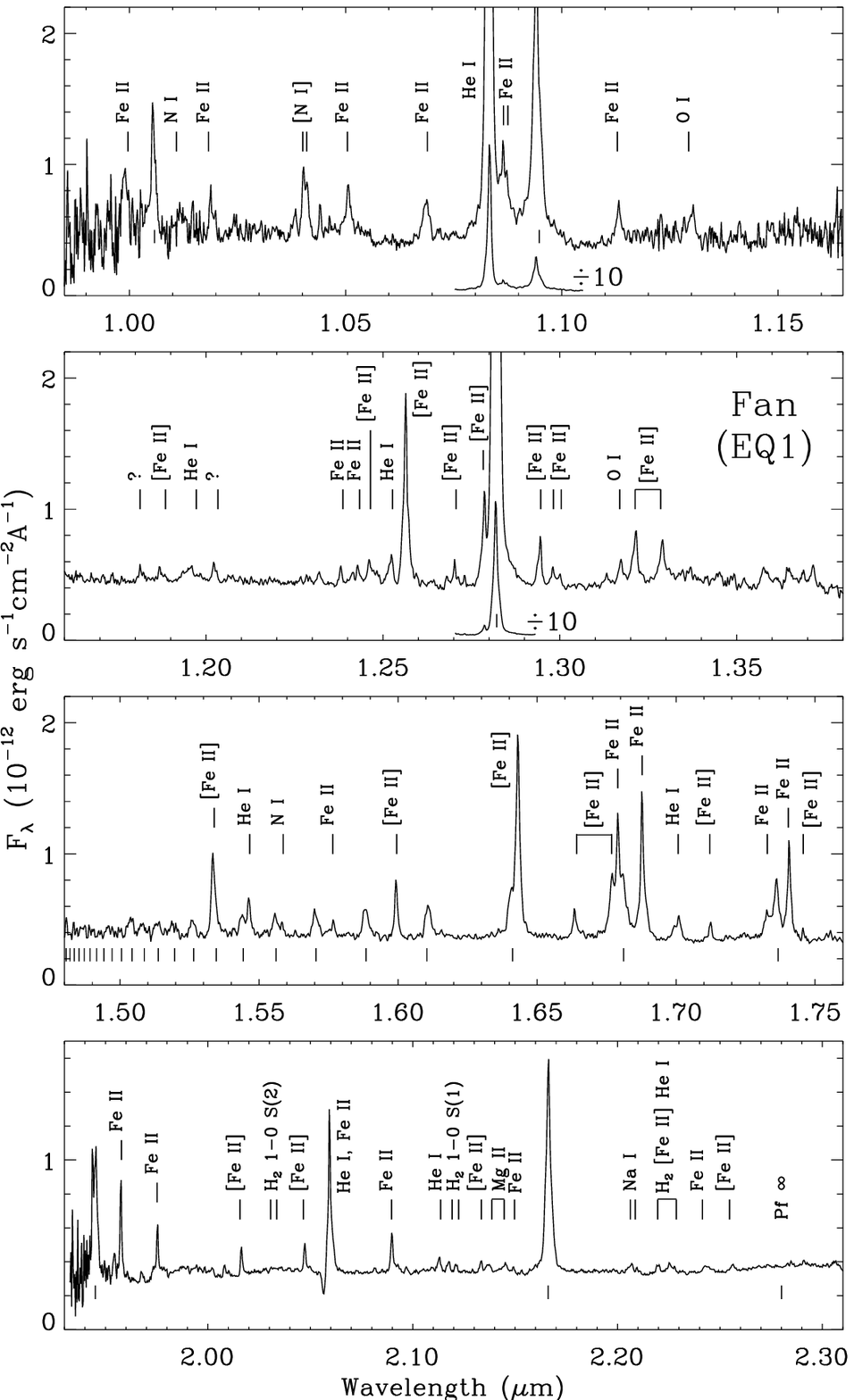,width=3.25in}\end{center}
\caption{Near-IR spectrum of the lower `Fan' extracted from a
1$\arcsec$ segment of the 0$\farcs$5-wide slit aperture, centered
2$\arcsec$ NW of the central star.  The long-slit aperture was
oriented at P.A.=45$\arcdeg$, roughly perpendicular to the major axis
of the Homunculus as shown by the dotted lines in Figure 1 (labeled as
position `EQ1').}
\end{figure}

\begin{figure}\begin{center}
\epsfig{file=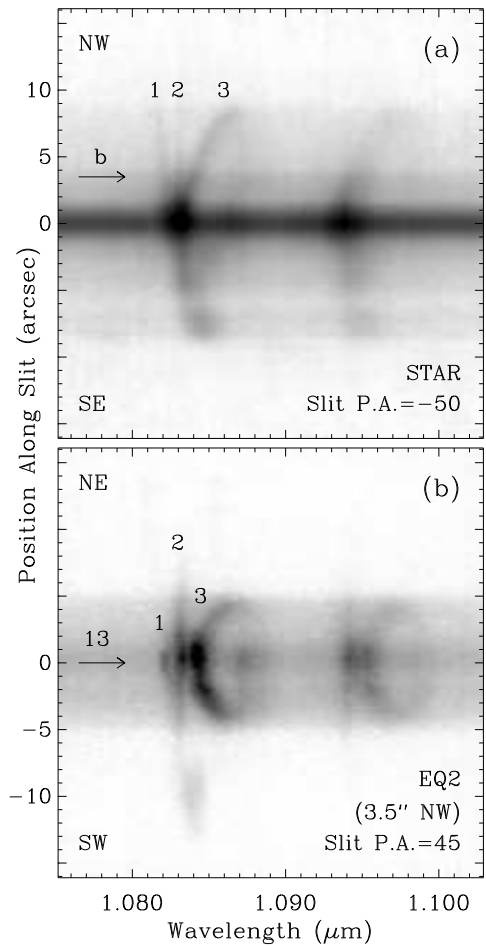,height=5in}\end{center}
\caption{Long-slit OSIRIS spectra showing He~{\sc i} $\lambda$10830 and
Pa$\gamma$.  Panel (a) is a segment of Figure 2$a$ with the slit
passing through the star at P.A.=-50$\arcdeg$.  The arrow identifies
the location 3$\farcs$5 NW of the star where this slit intersects that
used to obtain the long-slit spectrum shown in panel (b), obtained at
slit position `EQ2'.  The numbers 1 to 3 in both panels refer to
velocity components discussed in the text (\S 5) and shown in Figure
13.}
\end{figure}

\begin{figure}\begin{center}
\epsfig{file=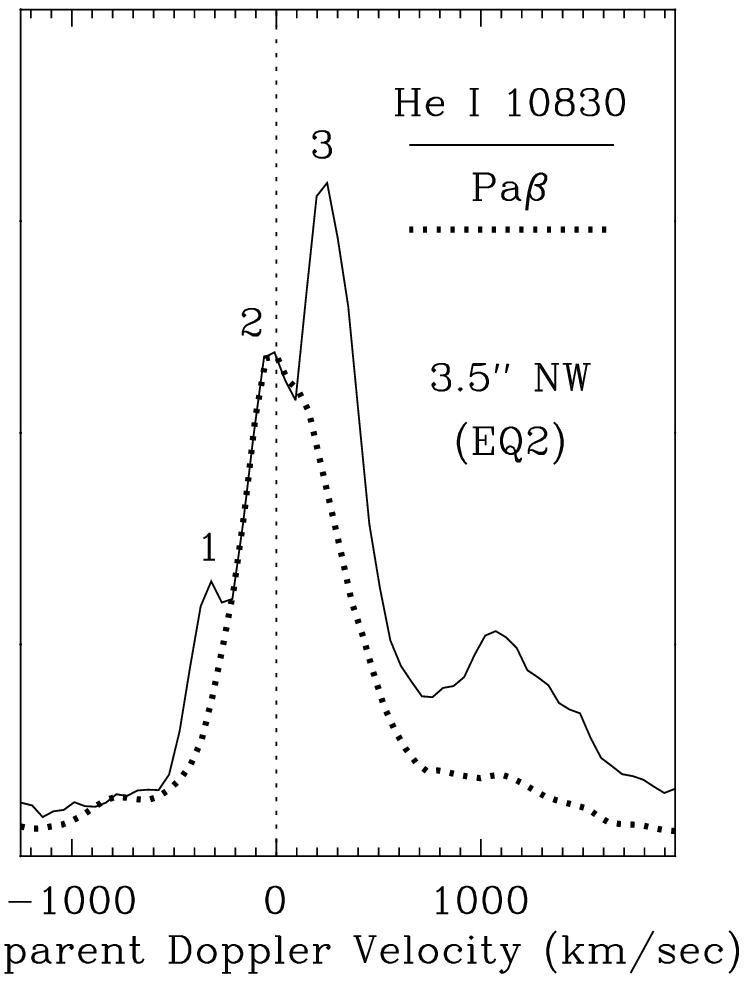,height=3in}\end{center}
\caption{Profile of He~{\sc i} $\lambda$10830 extracted from a
1$\arcsec$ segment of Figure 12$b$, at the position of the arrow
labeled `13' in that figure, with the slit at position `EQ2'.
Velocity components 1, 2, and 3 refer to features identified in Figure
12.  The red bump at $+$1100 km s$^{-1}$ is a combination of
redshifted He~{\sc i} emission and Fe~{\sc ii} $\lambda\lambda$10863,
10872.  The dotted line shows the profile of Pa$\beta$ observed at the
same position.}
\end{figure}

\subsection{IR Spectrum of the Fan}

Optical continuum images (Morse et al.\ 1998) show a prominent feature
extending a few arcseconds NW of the star, called the `Fan', and it
{\it looks} as if it is part of an equatorial spray.  However,
thermal-IR images reveal that the Fan is actually the dust
column-density minimum in the equatorial ejecta and it appears to be a
hole where we see through to the NW lobe (Smith et al.\ 1999, 2002).
This hole may also allow UV photons to escape to larger radii and
excite equatorial gas.  Part of the Fan shows blueshifted emission
with unusual excitation and strange kinematics (e.g., Zethson et al.\
1999; Johansson et al.\ 2000), as well as some emission lines like
[Sr~{\sc ii}] that are not seen anywhere else in the Homunculus
(Hartman et al.\ 2001; Zethson et al.\ 2001).

Figure 11 shows the near-IR spectrum at one position in the Fan
located $\sim$2$\arcsec$ NW of the central star.  At this position the
`star' and `EQ1' slits intersect (see Figure 1), and the spectrum in
Figure 11 was extracted from a 1$\arcsec$ segment of the EQ1 slit.
Like other positions in the Homunculus, strong [Fe~{\sc ii}] lines are
seen in the Fan.  [Fe~{\sc ii}] $\lambda$16435 is still one of the
strongest lines in the spectrum (perhaps due in part to contamination
from the LH), but the $\lambda$16435/Br$\gamma$ ratio here ($\sim$3)
is lower than in the polar lobes or LH.  The Br$\gamma$ flux in Table
2 corresponds to the narrow intrinsic emission in excess of the
reflected wind line; it is admittedly uncertain, but it seems
difficult to avoid the conclusion that direct radiative excitation
plays a more dominant role in the Fan than at other positions
discussed so far.  Other lines indicative of photoexcitation are also
enhanced in the Fan's spectrum, compared to the polar lobes or LH.

The [Fe~{\sc ii}] $\lambda$12567/$\lambda$16435 ratio is $\sim$1 in
the Fan.  This is higher than in the polar lobes or LH, and lower
reddening is consistent with the Fan's equatorial geometry in front of
the NW polar lobe.  The electron density derived from [Fe~{\sc ii}]
$a^4F-a^4D$ line ratios is $\gtrsim$10$^{5.3}$ cm$^{-3}$ (see Table
3), higher than in the LH or polar lobes, and near the critical
density.

However, the unique characteristic of the Fan, as compared to the LH
or polar lobes, is strong narrow emission from semi-forbidden Fe~{\sc
ii} lines in Figure 11.  Fe~{\sc ii} $\lambda$16787, $\lambda$16783,
$\lambda$17414, $\lambda$19746, $\lambda$20600 (blended with He~{\sc
i}), and $\lambda$20888 are all much stronger than in the polar lobes
or LH.  All these lines (see Table 2) are from the same multiplet
$z^4F-c^4F$, at higher energy than IR forbidden lines.  Interestingly,
the upper $c^4F$ term is also the lower term of the anomolous UV lines
near 2507 \AA, which may be excited by a fluorescent Ly$\alpha$
pumping mechanism (Johansson \& Letokhov 2001).  These infrared
Fe~{\sc ii} lines seen in the Fan are cascades from those anomolously
strong high-excitation fluorescent lines in the UV, and may be
variable as well.  The same IR lines are seen in the Weigelt blobs
(see below), but {\it not} in any polar ejecta. The Fan also glows in
He~{\sc i} $\lambda$10830, while the LH and polar lobes do not.  This
seems to confirm the conjecture from the $\lambda$16435/Br$\gamma$
ratio that photoexcitation dominates the Fan's spectrum, rather than
collisions.  More importantly, it appears that {\it the peculiar
fluorescent Fe~{\sc ii} lines are bright in equatorial gas, but are
absent or weak in polar ejecta}.

\subsection{An Extended Equatorial Disk?}

Near-IR spectra reveal some important clues about the kinematics of
$\eta$ Car's equatorial ejecta seen projected in front of the NW polar
lobe with blueshifted velocities.  The bright H$_2$ and [Fe~{\sc ii}]
emission lines discussed earlier did not show equatorial emission, but
Figure 12 shows equatorial gas in other lines, such as He~{\sc i}
$\lambda$10830.  Figure 12$a$ shows the kinematic structure of He~{\sc
i} $\lambda$10830 and Pa$\gamma$ along the slit oriented parallel to
the major axis of the Homunculus at P.A.=$-$50$\arcdeg$ (see Figure
1).  Toward the NW of the star, the He~{\sc i} line clearly branches
into three separate components, labeled 1, 2, and 3 in Figure 12$a$.
These components are also clearly seen and labeled in Figure 12$b$,
which shows a long-slit spectrum with the slit orthogonal to the polar
axis (slit position `EQ2' in Figure 1).  From Figures 12$a$ and $b$,
it is clear that component 3 corresponds to light from the star
reflected by expanding dust in the polar lobes, and will not be
discussed further.\footnotemark\footnotetext{However, note that
reflected He~{\sc i} $\lambda$10830 emission from the far wall of the
NW polar lobe affects Fe~{\sc ii} $\lambda\lambda$10863, 10872.}
Component 2 extends across and beyond the edges of the Homunculus in
Figures 12$a$ and $b$ at an almost constant low velocity.  It is not
seen toward the SE of the star; presumably component 2 resides near
the equator, and the far side is blocked by dust in the SE polar lobe.

Component 1 is only seen in He~{\sc i} $\lambda$10830.  It is narrow,
blueshifted, and has linearly increasing speed with distance (a Hubble
flow).  It most likely resides in the equatorial plane.  Component 1
extends to (and probably beyond) the edge of the NW polar lobe in
Figure 12$a$, but Figure 12$b$ reveals that it is confined to a narrow
range of azimuthal angles associated with the Fan.  Component 1
coincides spatially with the `purple haze' seen in {\it HST}/WFPC2
images (Morse et al.\ 1998). Since the Fan is probably caused by a
hole in the equatorial ejecta (Smith et al.\ 1999; 2002), it might
allow a beam of UV starlight to escape to large radii in a disk and
thereby excite the He~{\sc i} line.  Note that component 2 shows H
recombination emission, but component 1 does not.  Thus, components 1
and 2 might occupy separate regions of space along different
lines-of-sight to the central star.

\begin{figure*}\begin{center}
\epsfig{file=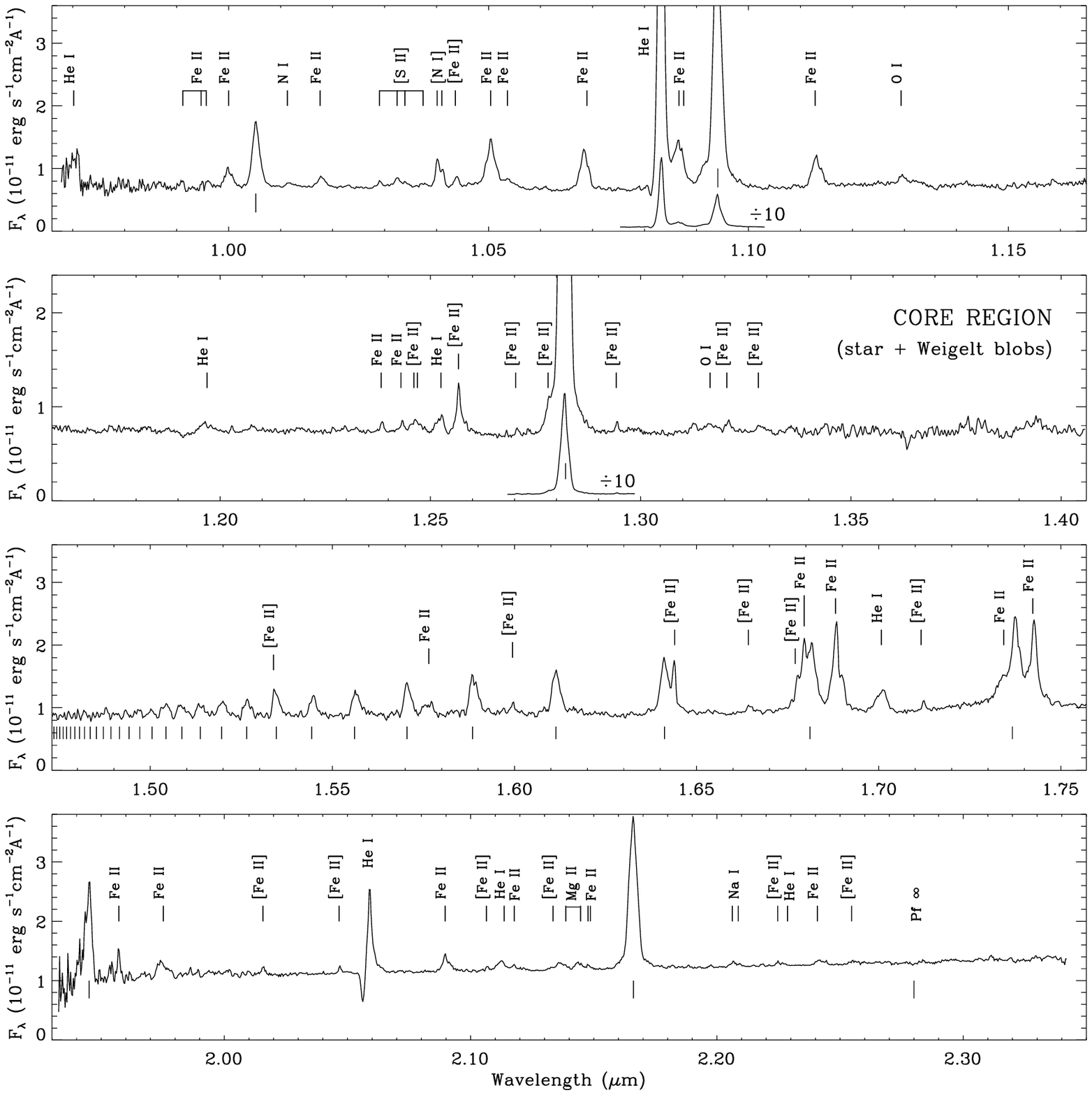,height=4.5in}\end{center}
\caption{Spectrum of the bright core region of the Homunculus,
extracted from a 0$\farcs$5 segment of the 0$\farcs$5-wide slit
aperture.  This includes the spectrum of the central star and Weigelt
blobs, but excludes emission from the Homunculus and Little
Homunculus.}
\end{figure*}

Figure 13 shows a tracing of the He~{\sc i} $\lambda$10830 line at
3$\farcs$5 NW of the star in the Fan.  Three velocity components are
quite distinct at this location, with velocities of $-$325, $-$35, and
$+$290 km s$^{-1}$ for components 1, 2, and 3, respectively.
Uncertainty in wavelength calibration is $\pm$20 km s$^{-1}$, but
relative velocities are reliable to within $\pm$5 km s$^{-1}$.

With a known inclination $i$=42$\fdg$5 derived in \S 3, the age $t$ in
years for {\it equatorial} gas with a given Doppler velocity $v$ (km
s$^{-1}$) can be inferred as a function of projected separation from
the star $r$ in arcsec (with linear motion) using

\begin{equation}
t = \frac{4.74 \ r \ D \ \tan i}{v}
\end{equation}

\noindent where $D$ is the heliocentric distance in pc.  Thus, at
3$\farcs$5 NW of the star, component 1 at $-$325 km s$^{-1}$ has a
probable age of only 105 $\pm$7 years --- perhaps ejected during the
minor eruption of $\eta$ Car in 1890.  If component 2 traces slow gas
near the equator, it may indicate older material with an age around
1000 years. Component 3 is not directly emitted by equatorial gas, as
noted above, so equation (3) is inapplicable.  Figure 13 shows that
component 1 is not seen in hydrogen lines, but Pa$\beta$ from
component 2 agrees quite well with He~{\sc i} emission.
Interestingly, neither He~{\sc i} $\lambda$10830 nor Paschen lines
show evidence for an equatorial component originating in the Great
Eruption.

\begin{figure*}\begin{center}
\epsfig{file=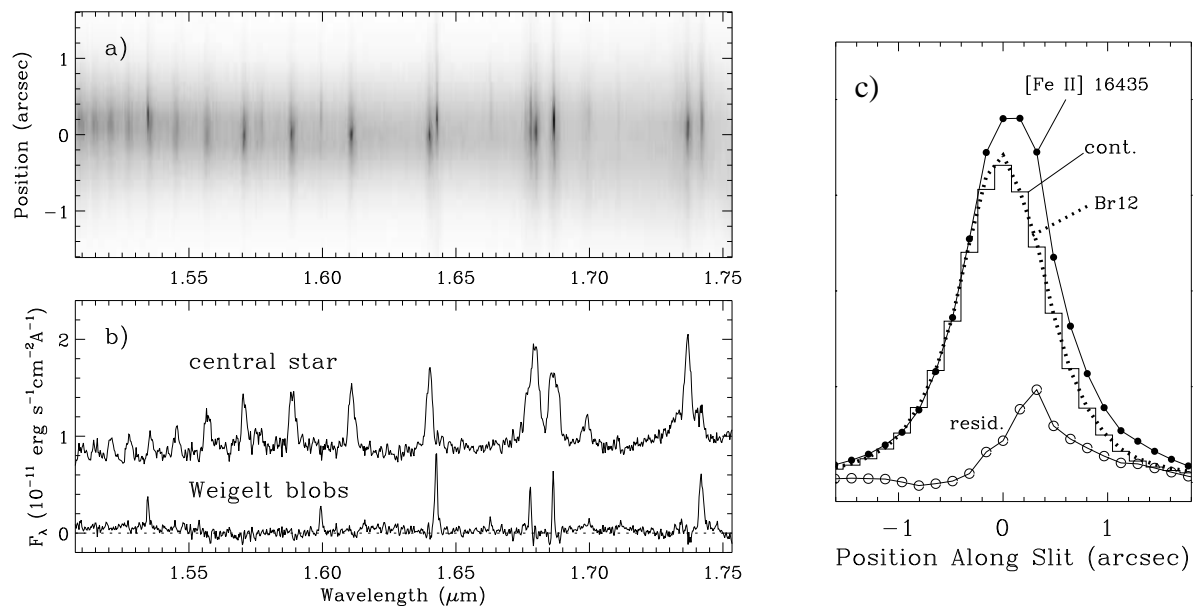,height=3.25in}\end{center}
\caption{Panel (a) shows a portion of Figure 2$e$ concentrating on the
inner few arseconds near the star.  Some narrow lines show real
positional offsets from the center.  0$\farcs$8-wide spectral
extractions to either side of zero were made.  Panel (b) shows
residual emission after subtracting scaled versions of these
extractions from one another, separating the spectra of the Weigelt
blobs and central star. (c) Spatial scans of [Fe~{\sc ii}]
$\lambda$16435 (filled circles), Br12 (dotted line), and adjacent
continuum (histogram).  Continuum emission was normalized to the Br12
emission.  Unfilled circles show residuals of subtracting the
continuum from [Fe~{\sc ii}] emission.}
\end{figure*}

These results might seem at odds with the conclusions of Zethson et
al.\ (1999), but the new inclination angle $i$=42.5 derived in \S 3
affects the strange radial velocities observed by those authors.  At a
particular position in the Fan at 1$\farcs$3 NE of the star (their
`Dsk 1'), Zethson et al.\ measured three velocity components at
$-$136, $-$81, and $-$42 km s$^{-1}$.  An incorrect value of
$i$$\approx$35$\arcdeg$ assumed at that time led them to infer ages of
160, 280, and 620 years, respectively.  However, with $i$=42$\fdg$5,
Zethson et al.'s velocity components at $-$136 and $-$81 km s$^{-1}$
would imply ejection during the 1890 outburst and the Great Eruption
of $\eta$ Car, respectively (see their Table 6).  Having derived a
similar inclination angle of $i$=41$\arcdeg$, Davidson et al.\ (2001)
found two velocity components in the Fan consistent with ejection
during the 1890 outburst and Great Eruption as well.  Thus, the new
inclination and structure of the Homunculus seem to partly resolve
some problematic kinematics of the equatorial ejecta, and also support
the presence of material younger than the Great Eruption.  So far,
this younger equatorial material from the 1890 outburst seems to be
seen only in emission lines, while proper motions with {\it HST}
images that trace primarily scattered continuum light do not detect it
(Morse et al.\ 2001).

\begin{figure}\begin{center}
\epsfig{file=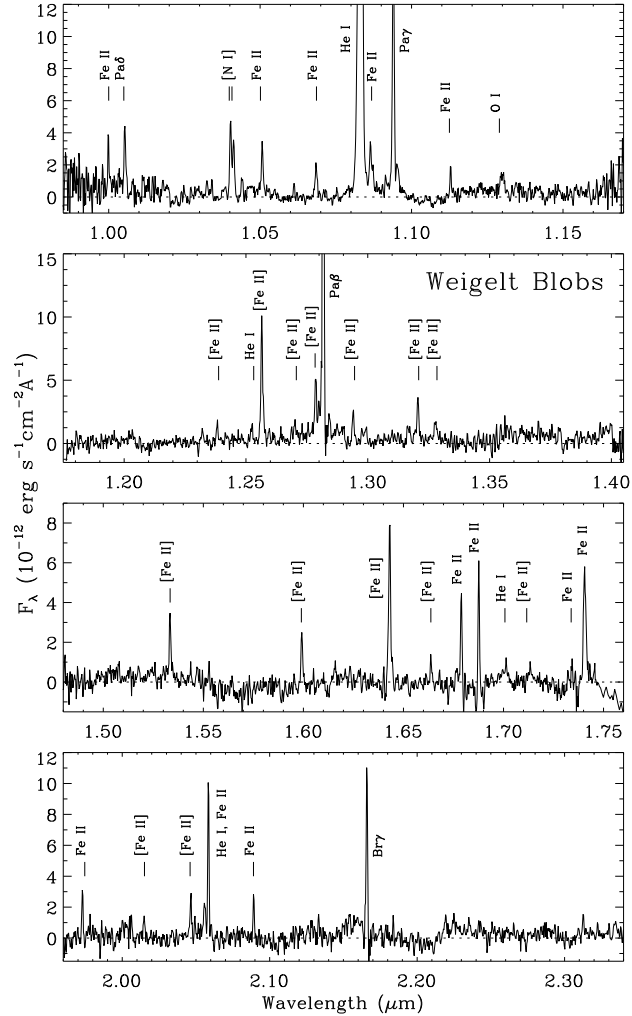,width=3.25in}\end{center}
\caption{Near-IR spectrum of the Weigelt blobs, made the same way as
described for Figure 15$b$.}
\end{figure}

\begin{table}
\scriptsize
\caption{IR Emission-lines in the Weigelt Blobs$^{a}$}
\begin{tabular}{llc} \hline\hline
Wavelength	&I.D.$^{b}$		&Flux		\\	
(\AA)		&		&(erg s$^{-1}$ cm$^{-2}$)		\\ \hline

9997		&Fe~{\sc ii} $(z^4F - b^4G)$	&2.84$\times$10$^{-11}$	\\
10049		&H~{\sc i} Pa$\delta$		&3.55$\times$10$^{-11}$	\\
10398		&[N~{\sc i}]			&3.37$\times$10$^{-11}$	\\
10407		&[N~{\sc i}]			&2.62$\times$10$^{-11}$	\\
10501		&Fe~{\sc ii} $(z^4F - b^4G)$	&3.06$\times$10$^{-11}$	\\
10608		&[Fe~{\sc iii}]			&5.59$\times$10$^{-12}$	\\
10686		&Fe~{\sc ii} $(^4P - ^4F)$	&2.08$\times$10$^{-11}$	\\
10830		&He~{\sc i}			&1.21$\times$10$^{-9}$	\\
10863,72	&Fe~{\sc ii} $(z^4F - b^4G)$	&3.42$\times$10$^{-11}$	\\
10912		&He~{\sc i}			&1.01$\times$10$^{-11}$	\\
10938		&H~{\sc i} Pa$\gamma$		&1.42$\times$10$^{-10}$	\\
11126		&Fe~{\sc ii} $(z^4F - b^4G)$	&1.27$\times$10$^{-11}$	\\
11287		&O~{\sc i} 			&1.82$\times$10$^{-11}$	\\
12384		&[Fe~{\sc ii}] $(z^6F - c^4F)$	&1.94$\times$10$^{-11}$	\\
12528		&He~{\sc i}			&1.77$\times$10$^{-11}$	\\
12567		&[Fe~{\sc ii}] $(a^6D - a^4D)$	&1.05$\times$10$^{-10}$	\\
12703		&[Fe~{\sc ii}] $(a^6D - a^4D)$	&2.57$\times$10$^{-11}$	\\
12788		&[Fe~{\sc ii}] $(a^6D - a^4D)$	&5.73$\times$10$^{-11}$	\\
12818		&H~{\sc i} Pa$\beta$		&3.45$\times$10$^{-10}$	\\
12943		&[Fe~{\sc ii}] $(a^6D - a^4D)$	&2.69$\times$10$^{-11}$	\\
12978		&[Fe~{\sc ii}] $(a^6D - a^4D)$	&2.37$\times$10$^{-11}$	\\
13205		&[Fe~{\sc ii}] $(a^6D - a^4D)$	&5.05$\times$10$^{-11}$	\\
13278		&[Fe~{\sc ii}] $(a^6D - a^4D)$	&3.07$\times$10$^{-11}$	\\
15335		&[Fe~{\sc ii}] $(a^4F - a^4D)$	&4.39$\times$10$^{-11}$	\\
15995		&[Fe~{\sc ii}] $(a^4F - a^4D)$	&3.39$\times$10$^{-11}$	\\
16435		&[Fe~{\sc ii}] $(a^4F - a^4D)$	&1.06$\times$10$^{-10}$	\\
16637		&[Fe~{\sc ii}] $(a^4F - a^4D)$	&1.91$\times$10$^{-11}$	\\
16787		&Fe~{\sc ii} $(z^4F - c^4F)$	&4.29$\times$10$^{-11}$	\\
16873		&Fe~{\sc ii} $(z^4F - c^4F)$	&5.74$\times$10$^{-11}$	\\
17002		&He~{\sc i}			&2.04$\times$10$^{-11}$	\\
17111		&[Fe~{\sc ii}] $(a^4F - a^4D)$	&1.55$\times$10$^{-11}$	\\
17338		&Fe~{\sc ii} $(z^4D - c^4P)$ ?	&1.31$\times$10$^{-11}$	\\
17414		&Fe~{\sc ii} $(z^4F - c^4F)$	&1.11$\times$10$^{-10}$	\\
17449		&[Fe~{\sc ii}] $(a^4F - a^4D)$	&8.29$\times$10$^{-12}$	\\
19746		&Fe~{\sc ii} $(z^4F - c^4F)$	&3.95$\times$10$^{-11}$	\\
20151		&[Fe~{\sc ii}] $(a^2G - a^2H)$	&2.14$\times$10$^{-11}$	\\
20460		&[Fe~{\sc ii}] $(a^4P - a^2P)$	&4.16$\times$10$^{-11}$	\\
20581,600	&He~{\sc i}, Fe~{\sc ii} $(z^4F - c^4F)$  &1.08$\times$10$^{-10}$	\\
20888		&Fe~{\sc ii} $(z^4F - c^4F)$	&3.32$\times$10$^{-11}$	\\
21655		&H~{\sc i} Br$\gamma$		&1.43$\times$10$^{-10}$	\\	\hline

\end{tabular}

$^{a}$See Figure 17. If more than one wavelength is given,
the line is \\ blended, and both lines are expected to contribute
significantly.

$^{b}$J subscripts for Fe$^{+}$ lines have been omitted.

\end{table}

\section{THE STAR'S SPECTRUM:  DIRECT AND SCATTERED LIGHT}

\subsection{The Central Core}

Figure 14 shows the flux-calibrated spectrum of the bright central
core of the Homunculus taken through a 0$\farcs$5$\times$0$\farcs$5
aperture in March 2001 during $\eta$ Car's high-excitation state. It
is essentially the combined spectrum of direct stellar light and the
nearby Weigelt blobs included in the aperture.  Figure 14 resembles
the IR spectrum with the same spectral resolution presented by Hamann
et al.\ (1994), who used a larger 4$\farcs$8$\times$4$\farcs$8
aperture.  Detailed differences exist between Hamann et al.'s spectrum
and Figure 14, but it is uncertain if they are due to temporal
variability or aperture-size effects.  Important changes have occurred
since 1990 when Hamann et al.\ obtained their spectrum, and a
$\sim$5$\arcsec$ aperture certainly includes extended emission.  The
contribution of compact ejecta and reflected light are examined below.

One detail worth mentioning here (with some bearing on $\eta$ Car's
variability) concerns how the strength of He~{\sc i} $\lambda$10830
varies with position or aperture size.  Measured in a small 0$\farcs$5
aperture, the bright central core region has a He~{\sc i}
$\lambda$10830 equivalent with of $\sim$205 \AA \ in Figure 14.
Farther out in the polar lobes, the He~{\sc i} $\lambda$10830
equivalent width is significantly less; 125 \AA \ in the NW lobe and
135 \AA \ in the SE lobe.  Independently, A.\ Damineli has been
monitoring $\eta$ Car's He~{\sc i} $\lambda$10830 line over the past
several years as seen in ground-based spectra with a larger aperture
(a few arcsec).  His measurements\footnotemark\footnotetext{\tt
http://www.igusp.usp.br/\~{ }damineli/etacar/index.html} give an
equivalent width roughly 15-20\% less than in Figure 14 at nearly the
same time.  Thus, the strength of this line is somewhat diluted in a
large aperture that includes more extended emission.

\subsection{Resolving the Spectrum of the Weigelt Blobs}

High-resolution imaging has resolved the bright core of
the Homunculus into a star plus three bright ejecta blobs
within 0$\farcs$3 (Weigelt \& Ebersberger 1986; Hofmann
\& Weigelt 1888; Weigelt et al.\ 1995; Morse et al.\ 1998).
These `Weigelt blobs' have a narrow emission-line spectrum different
from the central star, and probably reside near the equator (Davidson
et al.\ 1995; 1997). Emission from the Weigelt blobs is marginally
resolved in the near-IR spectra presented here, allowing their
combined spectrum to be separated from the central star for the first
time at IR wavelengths.  Figure 15$a$ shows a long-slit spectrum in
the H-band, concentrating on the inner few arceconds near the star. A
few narrow emission lines seem to be offset from the star's position.
Figure 15$c$ confirms this suspicion and proves that [Fe~{\sc ii}]
emission from the Weigelt blobs is marginally resolved.  Residual
emission after subtracting the continuum emission from [Fe~{\sc ii}]
has a centroid offset 0$\farcs$2 to 0$\farcs$4 from the star, as
expected.

Thus, it is possible to separate these blobs from the star by
extracting segments along the slit on either side of the star's
position.  To the NW the emission is dominated by the blobs, and to
the SE the star dominates.  Figure 15$b$ shows spectra of the Weigelt
blobs and star, after subtracting the adjacent
component.\footnotemark\footnotetext{The true spectrum of the Weigelt
blobs will include starlight scattered by dust in the blobs
themselves, but that has been subtracted here in addition to
instrumental scattering.}  The uncontaminated spectrum of the central
star in Figure 15$b$ shows only broad permitted lines from the stellar
wind, as expected following previous results with {\it HST} (Davidson
et al.\ 1995, 1997).  In general, Figure 15$b$ separates broad wind
lines from narrow nebular lines.  In some cases this applies to two
components of the same emission line; Fe~{\sc ii} $\lambda$16873 is a
salient example.

Figure 16 shows the full near-IR spectrum of the Weigelt blobs,
produced the same way as Figure 15$b$.  It differs from previously
published IR spectra of $\eta$ Car (Hamann et al.\ 1994), with only
narrow emission lines.  Table 4 lists fluxes that are not corrected
for extinction or reddening.  Infrared [Fe~{\sc ii}] line ratios
indicate an electron density in the Fe$^+$ zone above 10$^{5.6}$
cm$^{-3}$ (see Table 3).  This agrees with high densities inferred
from optical spectra (Hamann et al.\ 1999).

The [Fe~{\sc ii}] $\lambda$16435/Br$\gamma$ ratio is $\sim$1 for the
Weigelt blobs (Table 3).  These compact ejecta near the star are
almost certainly dominated by radiative excitation, so this reaffirms
comments made earlier that the Fan is also dominated by
photoexcitation.  Like the Fan, and unlike the polar lobes and LH, the
Weigelt blobs show bright Fe~{\sc ii} $z^4F-c^4F$ transitions that are
cascades from levels populated by Ly$\alpha$ fluorescence, directly
following the bright lines near 2507 \AA \ (Johansson \& Letokhov
2001).  As noted earlier, it appears that these fluorescence lines (at
least in the IR) are only enhanced in equatorial ejecta. More
generally, IR spectra of the Weigelt blobs, Fan, LH, and the polar
lobes of the Homunculus all suggest that {\it gas near the equator is
radiatively excited, while polar directions are dominated by
collisional excitation.}  Shock excitation in polar ejecta is
understandable if the central star has a slow equatorial wind, and a
fast polar wind (Smith et al.\ 2003) that is catching-up with
previously ejected material.  Excitation by radiative heating may also
be possible in the polar lobes (Ferland et al.\ 2002). Strong
photoexcitation and peculiar Ly$\alpha$ fluorescence mechanisms
localized in the equator seem harder to understand, but qualitatively
they suggest that Lyman continuum radiation escapes preferentially at
low latitudes in $\eta$ Car's otherwise dense wind.  This is broadly
consistent with ideas about density structure in the wind expressed by
Smith et al.\ (2003). This directional dependence of excitation in the
Homunculus places interesting constraints on the geometry of the UV
radiation field from the central star, expected to vary during $\eta$
Car's spectroscopic cycle.

\begin{figure*}\begin{center}
\epsfig{file=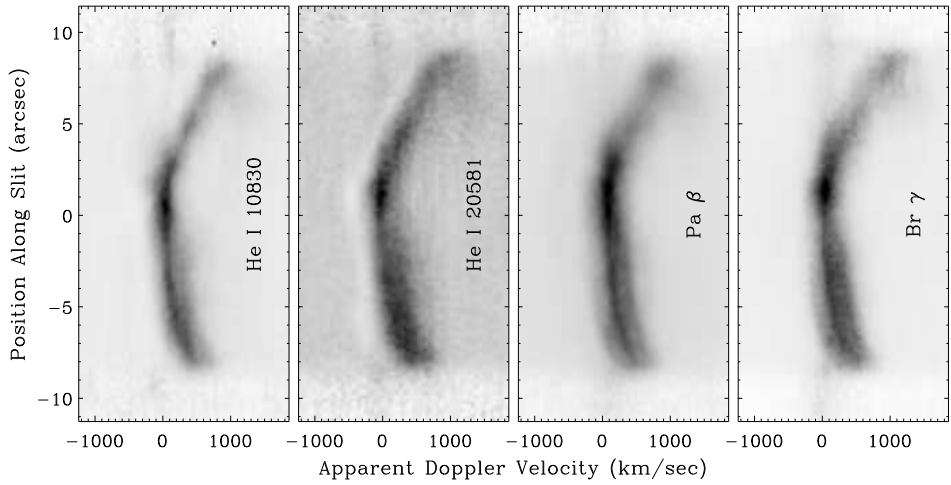,height=2.5in}\end{center}
\caption{Long-slit spectra for reflected stellar-wind emission
lines as a function of position in the Homunculus, corresponding to
the NE slit position (see Figure 1).  As previously, negative offsets
are toward the SE, and positive offsets are NW along the slit.}
\end{figure*}
\begin{figure*}\begin{center}
\epsfig{file=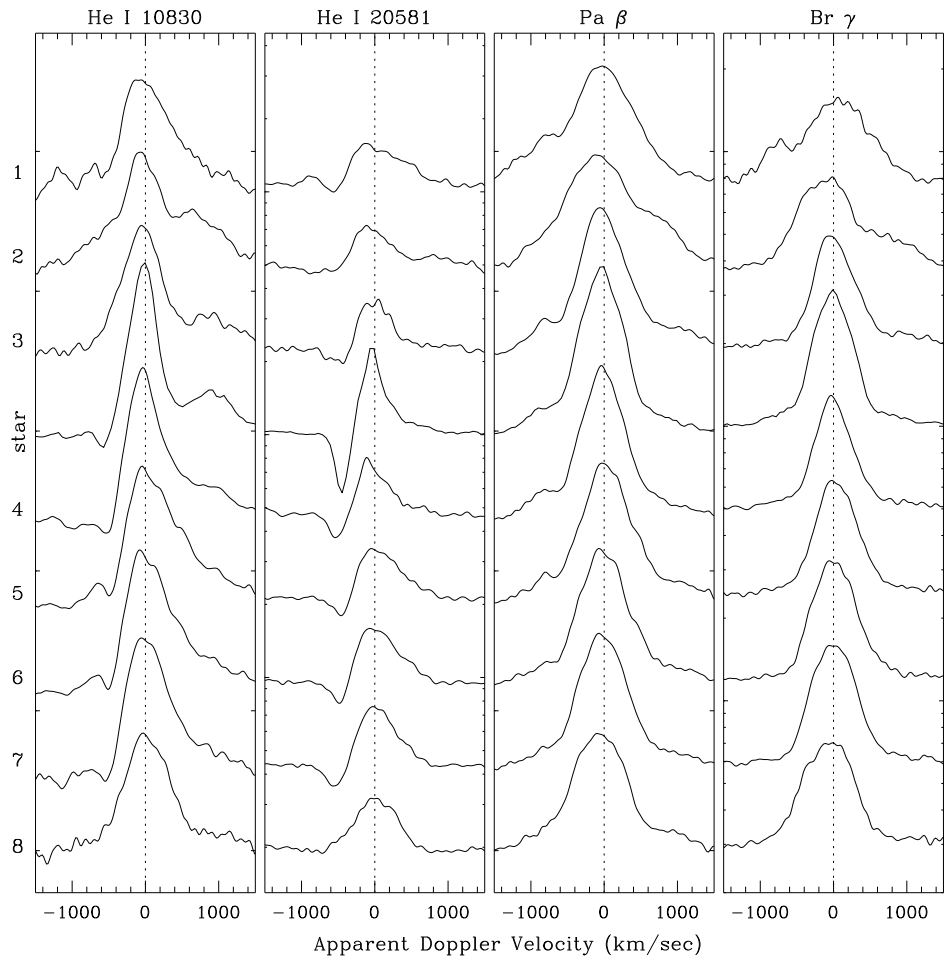,height=5in}\end{center}
\caption{Tracings of line profiles for He~{\sc i} $\lambda$10830,
$\lambda$20581, Pa$\beta$, and Br$\gamma$ from Figure 17 at various
positions along the slit (labeled at left; see Table 5).  Line
profiles of the bright central core (star $+$ Weigelt blobs) from
Figure 14 are also included.}
\end{figure*}

\subsection{Reflected Spectra in the Polar Lobes \\ and the Stellar Wind Geometry}

Despite the prominent H$_2$ and [Fe~{\sc ii}] emission described
above, the Homunculus is primarily a dusty reflection nebula.  Thus,
it provides the rare opportunity to observe the scattered spectrum of
a star from multiple directions.  Hillier \& Allen (1992), Hamann et
al.\ (1994), and others have shown that the spectrum observed in the
middle of the SE lobe is essentially a reflected spectrum of the
central star, with less contamination from narrow emission lines
arising in circumstellar gas.  However, so far reflected spectra have
only been published for wavelengths below 1 $\micron$.

Figure 17 shows details of the velocity structure at the NE slit
position for four reflected stellar-wind emission lines: He~{\sc i}
$\lambda$10830, He~{\sc i} $\lambda$20581, Pa$\beta$, and Br$\gamma$,
and Figure 18 shows tracings of the velocity profiles for these lines
at several different offset positions listed in Table 5.
Representative latitudes for these reflected spectra are also listed
in Table 5, derived from the structure of the polar lobes in \S 3.
Figure 19 shows line profiles for the central star and reflected light
at a few representative positions in the Homunculus, but normalized to
the same continuum level and superimposed on one another.  The
position 1$\arcsec$ NE of the star is chosen to show reflected light
from the star at nearly the same latitude as our direct line-of-sight,
and the positions in the SE and NW polar lobes are chosen to represent
reflected light as seen from high and low latitudes in the stellar
wind, respectively.

{\it Hydrogen lines (Pa$\beta$ and Br$\gamma$).}  Near-IR hydrogen
lines show nearly symmetric, pure emission profiles at all latitdues,
with minor variations in line shapes.  Balmer lines, on the other
hand, exhibit strong latitudinal dependence of P Cygni absorption
(Smith et al.\ 2003). With the spectral resolution used here, near-IR
hydrogen lines cannot offer definite constraints on the wind geometry,
but they support the hypothesis that asymmetry seen in Balmer
absorption (formed farther out in the wind) is due to an
ionization/recombination imbalance caused by a density enhancement in
the polar wind (see Smith et al.\ 2003).

\begin{table}
\caption{Reflected Line-Profile Extractions$^{a}$}
\begin{tabular}{llcc} \hline\hline
Position 	&Position	 	&$\Delta$V 	&Latitude$^{b}$	\\
		&(arcsec)		&(km s$^{-1}$)	& (deg)				\\ \hline

1	&NW 7.5		&830		&58	\\
2	&NW 5.5		&480		&47	\\
3	&NW 2.6		&165		&26	\\
STAR	&0		&0		&48	\\
4	&NE 1.0		&75		&45	\\
5	&SE 2.4		&120		&63	\\
6	&SE 4.0		&190		&72	\\
7	&SE 5.7		&240		&83	\\
8	&SE 8.0		&430		&71	\\ \hline

\end{tabular}

$^a$See Figure 18.

$^b$Uncertainty in latitude is at least $\pm$5$\arcdeg$.

\end{table}

{\it Helium lines.}  Unlike He~{\sc i} recombination lines at optical
wavelengths, He~{\sc i} lines at 10830 and 20581 \AA \ show P Cygni
absorption at all viewing angles except some positions in the NW lobe
contaminated by equatorial emission (see \S 5.2).  Tracings of
$\lambda$20581 in Figures 18 and 19 show that the star has much deeper
absorption than any reflected positions, even positions at roughly the
same latitude; perhaps this is a radiative transfer effect similar to
the excess emission seen in direct light from the central star at both
optical and IR wavelengths (see Smith et al.\ 2003 and below).
Tracings of reflected He~{\sc i} $\lambda$20581 show weak latitudinal
dependence in P Cyg absorption at about $-$550 km s$^{-1}$, but the
He~{\sc i} $\lambda$10830 absorption structure seen in the long-slit
spectrum in Figure 20 is quite remarkable.  The complex blueshifted
He~{\sc i} $\lambda$10830 absorption in Figure 20 may be a combination
of P Cygni absorption in the stellar wind at velocities of roughly
$-$550 km s$^{-1}$, plus nebular absorption from a shell outside the
Homunculus, analogous to the Ca~{\sc ii} absorption mentioned earlier.
The fast blueshifted absorption has a kinematic structure and radial
velocity similar to the outer bubble seen in [Fe~{\sc ii}]
$\lambda$16435 in Figures 4 and 5.  Thus, this outer nebular material
may be responsible for the high-speed blueshifted absorption in the
$\lambda$10830 line seen by Damineli et al.\ (1998), rather than a
fast component of the stellar wind.  It may also causes the fast
blueshifted absorption component seen in some UV resonance lines by
Viotti et al.\ (1989).  There is fast ($\sim$1000 km s$^{-1}$)
material in $\eta$ Car's wind, but so far it has only been detected in
reflected Balmer lines (Smith et al.\ 2003).

\begin{figure}\begin{center}
\epsfig{file=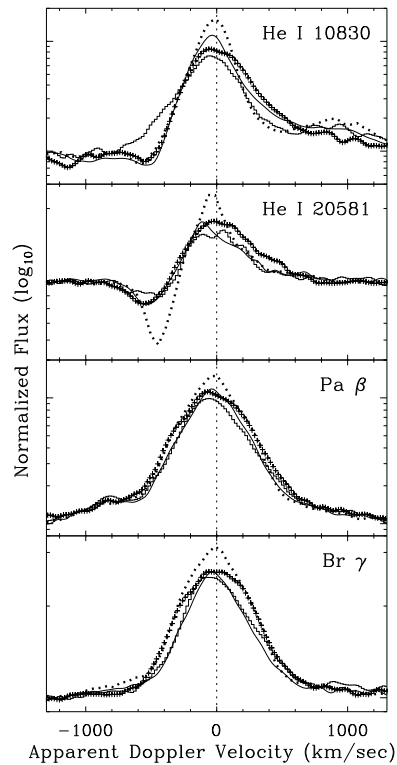,height=4in}\end{center}
\caption{Selected tracings from Figure 18 for direct light from the
central star (dotted line), overplotted with profiles seen in
reflected light from the SE lobe at position 7 (crosses), the NW lobe
at position 3 (histogram), and 1$\arcsec$NE of the star at position 4
(thin solid line).  See Table 5 for more details.}
\end{figure}

\begin{figure}\begin{center}
\epsfig{file=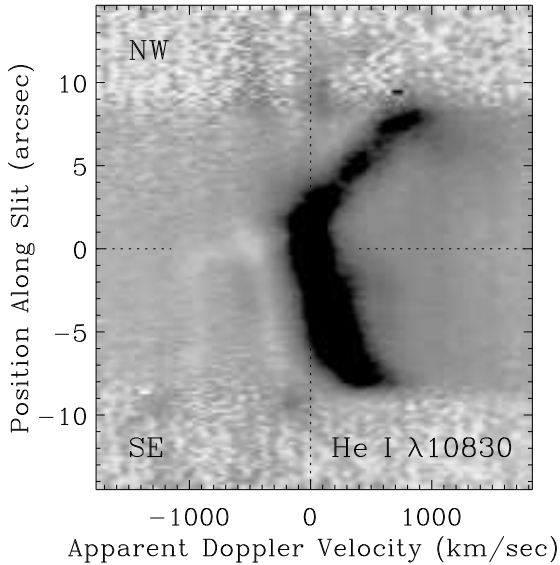,width=2.75in}\end{center}
\caption{Same as for He~{\sc i} $\lambda$10830 in Figure 17, but
stretched horizontally and displayed to emphasize faint emission.
Notice the blueshifted absorption in the SE lobe.}
\end{figure}

Reflected emission peaks for both H and He~{\sc i} lines have
comparable strengths at various latitudes, and all are weaker than the
peak emission for lines seen in direct light from the central star.
This effect is significant -- line profiles in Figure 19 are plotted
with a {\it logarithmic} vertical axis.  The same effect has been
noted for optical lines in $\eta$ Car, and does not yet have a
suitable explanation (Smith et al.\ 2003).

Another trait shown by both He~{\sc i} and H line profiles in Figure
19 is that reflected lines seen in the SE polar lobe (crosses) are
always broader and extend farther toward the red than reflected lines
seen in the NW polar lobe (histogram).  This is also apparent in
Figures 17 and 20, and is consistent with the behavior of optical
He~{\sc i} lines and the geometry they imply (Smith et al.\ 2003);
Balmer emission lines do not show this trend.  Thus, infrared He~{\sc
i} and H lines may trace a deep and important region of the wind
(perhaps 10 to 100 stellar radii; see Hillier et al.\ 2001) where
asymmetries arise.  Spatial variation in these line profiles are worth
a more detailed look with higher spectral resolution.

\section{SUMMARY}

Spatially-resolved IR data reveal numerous details of the apparent
spectrum as a function of position in the Homunculus, described above.
General results are summarized here:

1.  Kinematic structure of [Fe~{\sc ii}] and H$_2$ lines in the
Homunculus give the clearest picture yet of the geometry of the polar
lobes, with $i$$\approx$42$\fdg$5.  The far side of the SE lobe is
seen clearly for the first time in these emission lines.

2.  These near-IR spectra reveal bright [Fe~{\sc ii}] emission from a
`Little Homunculus' that was discovered previously at optical
wavelengths (Ishibashi et al.\ 2003), and offer some constraints on
the physical conditions in that relatively small bipolar nebula.  The
Little Homunculus has no H$_2$ emission.

3.  [Fe~{\sc ii}] emission and He~{\sc i} $\lambda$10830 absorption
reveal the existence of fast moving material outside the Homunculus
that is projected along the line-of-sight to the SE polar lobe.

4.  The Fan and other equatorial ejecta have different spectral
characteristics than the polar lobes, including blueshifted He~{\sc i}
$\lambda$10830 emission that apparently comes from material ejected
after the Great Eruption.

5.  The Weigelt blobs appear to be marginally resolved in these
long-slit near-IR spectra.  Subtracting the star's spectrum allows the
IR spectrum of the Weigelt blobs to be isolated for the first time.

6.  There is a strong {\it directional dependence of excitation} in
$\eta$ Car's circumstellar ejecta.  IR spectra of equatorial gas are
characteristic of radiative excitation and Ly$\alpha$ fluorescence,
whereas polar ejecta are collisionally excited (either due to shocks
or indirect radiative heating).  In other words, ejecta near the
equator ``see'' strong emission in the Lyman continuum, and polar
ejecta do not.  This has important implications for the geometry of
$\eta$ Car's UV radiation field, and that geometry seems qualitatively
consistent with the stellar wind structure during $\eta$ Car's normal
high-excitation state proposed by Smith et al.\ (2003).

7.  Reflected emission in the Homunculus provides multiple viewing
angles to the star, and IR wind lines show some interesting variation
(although not as dramatic as Balmer absorption).  Both He~{\sc i}
$\lambda$10830 and $\lambda$20581 show P Cygni absorption at nearly
all latitudes at roughly $-$550 km s$^{-1}$, whereas IR hydrogen lines
show pure emission profiles.  Other details are discussed as well.
For instance, high speed He~{\sc i} $\lambda$10830 absorption arises
in an outer nebula, rather than in a fast component of the stellar
wind.

As implied frequently in this paper, these results can be improved
upon with higher spectral and spatial resolution that will soon become
available.  In particular, such data can give us the first look at the
detailed clumpy structure of H$_2$ in the polar lobes and its relation
to ionized gas, they can give us the best kinematic map of the Little
Homunculus, and adaptive optics will hopefully allow the IR spectrum
of the Weigelt blobs to be spatially-resolved without uncertainties
associated with the method used in this paper.  Improved spectral
resolution combined with long-slit spectroscopy will also help us
understand latitudinal variations in He~{\sc i} $\lambda$10830 and
other reflected lines.  Finally, we must not forget that $\eta$ Car is
a notorious variable star.  The directional dependence of excitation
indicated by IR emission lines may respond to changes in the
latitudinal density structure of $\eta$ Car's wind during its 5.5 year
spectroscopic cycle (Smith et al.\ 2003), so emission from
circumstellar gas should also be monitored with long-slit IR
spectroscopy. \\

\smallskip\smallskip\smallskip\smallskip
\noindent {\bf ACKNOWLEDGMENTS}
\smallskip
\footnotesize

\noindent I am grateful to Bob Blum and Patrice Bouchet for their help
during the OSIRIS observing run, and to an anonymous referee for
useful comments on the manuscript. Examining data obtained with {\it
HST}/STIS has colored my interpretation of the IR spectra, and STIS
measurements included in Figure 5 resulted from extant work done in
collaboration with K.\ Davidson, T.R.\ Gull, K.\ Ishibashi, and J.\
Hillier.  Conversations with K.\ Ishibashi were helpful in regard to
the Little Homunculus discussed in \S 4, and I benefitted from
discussions with G.\ Ferland about excitation in the Homunculus.  NOAO
payed for my travel to Chile and accomodations while at CTIO.  I am
also grateful for the support of a NASA GSRP fellowship from Goddard
Space Flight Center.

\label{lastpage}
\end{document}